\AtBeginDocument{}
\documentclass[aps,prl,twocolumn,superscriptaddress,calc,epsfig,10pt,floatfix]{revtex4-1}

\usepackage{graphicx}
\usepackage{graphics}
\usepackage{amssymb,amsmath}
\usepackage{wasysym}
\usepackage{float}
\usepackage{physics}
\usepackage{color}
\usepackage[normalem]{ulem} 
\usepackage[pdftex,breaklinks,colorlinks=true,linkcolor=blue,citecolor=blue,urlcolor=blue]{hyperref}
\usepackage{tikz}
\usetikzlibrary{calc,shapes}
\usepackage{lineno}  

\newcommand{\mitCUAaddress}{Department of Physics, MIT-Harvard Center for Ultracold Atoms, and Research Laboratory of Electronics, MIT, Cambridge, Massachusetts 02139, USA}

\begin{document}

\title{Direct observation of non-local fermion pairing in an attractive Fermi-Hubbard gas}

\author{Thomas Hartke, Botond Oreg, Carter Turnbaugh, Ningyuan Jia, and Martin Zwierlein}
\date{\today}
\affiliation{\mitCUAaddress}

\begin{abstract}
Pairing of fermions lies at the heart of superconductivity, the hierarchy of nuclear binding energies and superfluidity of neutron stars.
The Hubbard model of attractively interacting fermions provides a paradigmatic setting for fermion pairing, featuring a crossover between Bose-Einstein condensation (BEC) of tightly bound pairs and Bardeen-Cooper-Schrieffer (BCS) superfluidity of long-range Cooper pairs, and a ``pseudo-gap'' region where pairs form already above the superfluid critical temperature.
We here directly observe the non-local nature of fermion pairing in a Hubbard lattice gas, employing spin- and density-resolved imaging of $\sim$1000 fermionic ${}^{40}$K atoms under a bilayer microscope. Complete fermion pairing is revealed by the vanishing of global spin fluctuations with increasing attraction.
In the strongly correlated regime, the fermion pair size is found to be on the order of the average interparticle spacing.
We resolve polaronic correlations around individual spins, resulting from the interplay of non-local pair fluctuations and charge-density-wave order.
Our techniques open the door toward in-situ observation of fermionic superfluids in a Hubbard lattice gas.
\end{abstract}

\maketitle

Long-range Cooper pairs form in a Fermi gas for even the weakest attraction between fermions. With increasing interaction, fermion pairs become more tightly bound, as the system undergoes a smooth crossover from BCS superfluidity towards a BEC of molecular pairs~\cite{Inguscio2008Ultracold,Zwerger2012BCS, Randeria2014Crossover}.
In the BCS limit, pair formation and the onset of superfluidity occur at the same temperature, but in the crossover pairs are expected to form already at temperatures $T^*$ above the critical temperature $T_c$ for superfluidity. In this so-called ``pseudo-gap'' regime the pair size should be on the order of the interparticle spacing and pairing strongly affected by many-body effects~\cite{Randeria1992Pairing,Trivedi1995Deviations}. The character of this strongly correlated regime, situated between a Fermi liquid and a normal Bose liquid, is a matter of debate, whose resolution should impact understanding of other strongly coupled fermion systems, such as the high-$T_c$ cuprates and twisted bilayer graphene~\cite{Lee2006Doping,Cao2018Unconventional,Chen2005BCS}. 
The rich physics of the BEC-BCS crossover is captured by the attractive Fermi-Hubbard model, a spin-1/2 gas of fermions hopping on a lattice with on-site interactions between unlike spins~\cite{Scalettar1989Phase,Singer1998On,Paiva2004Critical,Paiva2010Fermions,Fontenele2022Two,Singer1996From,Moreo1991Two,Keller1999Thermodynamics,Bauer2009Dynamical}. Through a particle-hole transformation it stands in one-to-one correspondence with the repulsive Hubbard model~\cite{Ho2009Quantum,Gall2020Simulating}, believed to hold the key towards understanding high-temperature superconductivity. 
The model can be realized using neutral fermionic atoms in optical lattices with tunable interactions. 
Recent investigations have found spectral gaps~\cite{Brown2020Angle}, correlations between local pairs~\cite{Mitra2018Quantum}, and evidence for inter-spin correlations from density profiles~\cite{Chan2020Pair}. 


\begin{figure}[!t]
	\centering
	\includegraphics[width=\columnwidth]{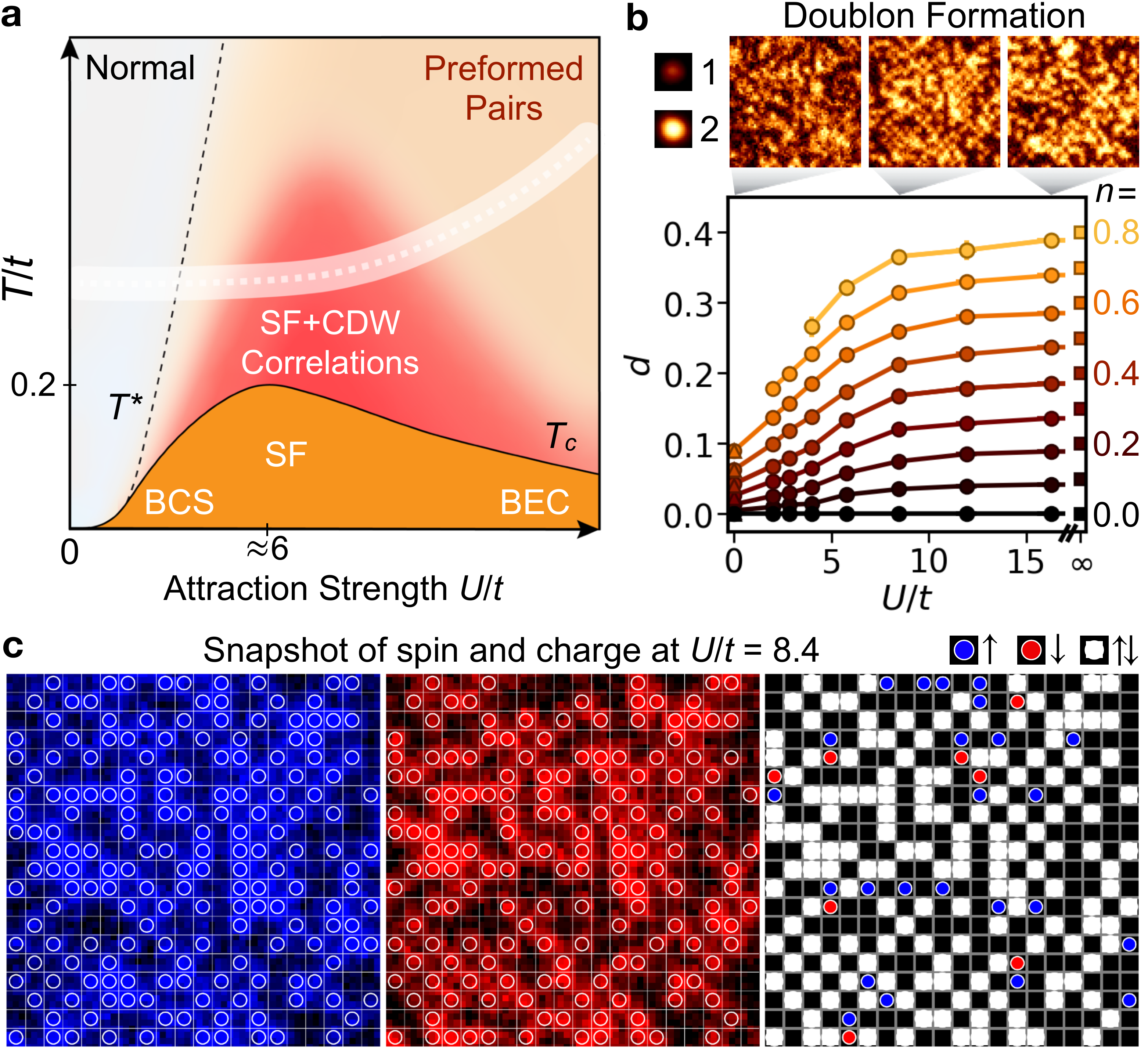}
	\caption{\textbf{Atom-resolved detection of an attractive Fermi-Hubbard gas.} 
	(a)~Qualitative phase diagram of the attractive Fermi-Hubbard model vs.~onsite attraction $U/t$ and temperature $T/t$ at density $n{\approx}0.8$~\cite{Scalettar1989Phase,Singer1998On,Paiva2004Critical,Paiva2010Fermions,Fontenele2022Two}.
	Below a critical temperature $T_c$, attractive fermions form a BCS or BEC superfluid~(SF).
	In the pseudo-gap regime between $T_c$ and pairing temperature $T^*$, accessed in this work~(white shading), increasing attraction drives pair formation, with pairs exhibiting charge-density-wave~(CDW) and superfluid correlations. 
    (b)~Measured doublon density $d$~(circles) at fixed density $n$ vs.~$U/t$, from the non-interacting limit $d{=}(n/2)^2$~(triangles) to the fully-paired limit $d{=}n/2$~(squares), with representative images of the full density in ${\sim}20{\times} 20$ site regions shown above. 
    (c)~Snapshot of full spin-and-density readout of a strongly-correlated gas at $U/t {=} 8.4(4)$ and $T/t {=} 0.36(5)$. The spin up~(blue), spin down~(red), and combined images~(right side) are obtained via bilayer quantum gas microscopy~\cite{SI,Hartke2020Doublon}.
	}
	\label{fig:Fig_AttractivePhaseDiagram}
\end{figure}

In this work we observe the formation and spatial ordering of non-local fermion pairs in the pseudo-gap regime of an attractive Hubbard gas confined to two dimensions. We employ bilayer quantum gas microscopy to detect the in-situ location and spin of each fermion in every experimental shot~\cite{SI, Koepsell2020Robust,Hartke2020Doublon}. 
Access to microscopic spin and density correlations reveals the formation of non-local pairs, the development of long-range spatial correlations between pairs, and the interplay of pair fluctuations with this density-wave order. 

The phase diagram of the attractive Fermi-Hubbard model is shown in Fig.~\ref{fig:Fig_AttractivePhaseDiagram}(a) as a function of the attractive onsite interaction strength $U$, tunneling amplitude $t$, and temperature $T$~\cite{Scalettar1989Phase,Singer1998On,Paiva2004Critical,Paiva2010Fermions,Fontenele2022Two}. 
For weak attraction $U{\ll} t$ a BCS superfluid of long-range fermion pairs forms with $T_c {=} T^*$, reflecting the exponentially weak pair binding. In the opposite limit of strong attraction $U{\gg} t$, all fermions are bound into local onsite pairs below a dissociation temperature $T^*{\sim} U$. These pairs condense at the critical temperature of Bose-Einstein condensation $T_c$, proportional to the pair density $n_p$ and pair tunneling rate $t_p{\sim} t^2/U$.
A peak of the condensation temperature $T_c/t {\approx} 0.2$ is expected to occur at $U/t{\approx} 6$ and density~$n{\approx}0.8$~\cite{Scalettar1989Phase,Singer1998On,Paiva2004Critical,Paiva2010Fermions,Fontenele2022Two}.   
Above the transition temperature, superfluid correlations compete with the formation of a checkerboard charge-density-wave~\cite{Mitra2018Quantum}. At half filling~(density~$n{=}1$) this competition persists down to $T{=}0$ and prevents condensation. In this work we employ a filling $n{\approx}0.8$, staying in a regime where the ground state is a paired superfluid~\cite{Scalettar1989Phase, Moreo1991Two}. 

As a first measure of strong pairing in the attractive Hubbard gas, we measure the doublon density $d$ for increasing interaction strength $U/t$ across the phase diagram in Fig.~\ref{fig:Fig_AttractivePhaseDiagram}(a). 
At fixed density, $d$ increases from the non-interacting limit $d{=}(n/2)^2$ of random encounters of unlike spins to the fully-paired limit $d{=}n/2$~(Fig.~\ref{fig:Fig_AttractivePhaseDiagram}(b))~\cite{Bauer2009Dynamical}. 
At intermediate attraction, strong checkerboard ordering of doublons is observed, shown in Fig.~\ref{fig:Fig_AttractivePhaseDiagram}(c) at $U/t {=} 8.4(4)$ and $T/t {=} 0.36(5)$. 

Multiple neighboring sites containing a single spin up and spin down are present among doublons in Fig.~\ref{fig:Fig_AttractivePhaseDiagram}(c). These correlated pairs of single spins are evidence of the non-local nature of fermion pairs. The microscopic mechanism is the virtual dissociation of a doublon into spatially separate pairing partners, with matrix element $t$ and intermediate energy cost $U$, which perturbatively lowers the energy of a pair by $4t^2/U$. 
Because pairs are composed of fermions, dissociation can only occur if a nearby site does not already contain a like spin. This leads to effective nearest-neighbor repulsive interactions between pairs~\cite{SI} which in turn are the source of long-range charge-density-wave (CDW) order. The presence of these delocalized pairs also demonstrates that the doublon density $d$ is an incomplete measure of pairing.

\begin{figure}[!t]
	\centering
	\includegraphics[width=\columnwidth]{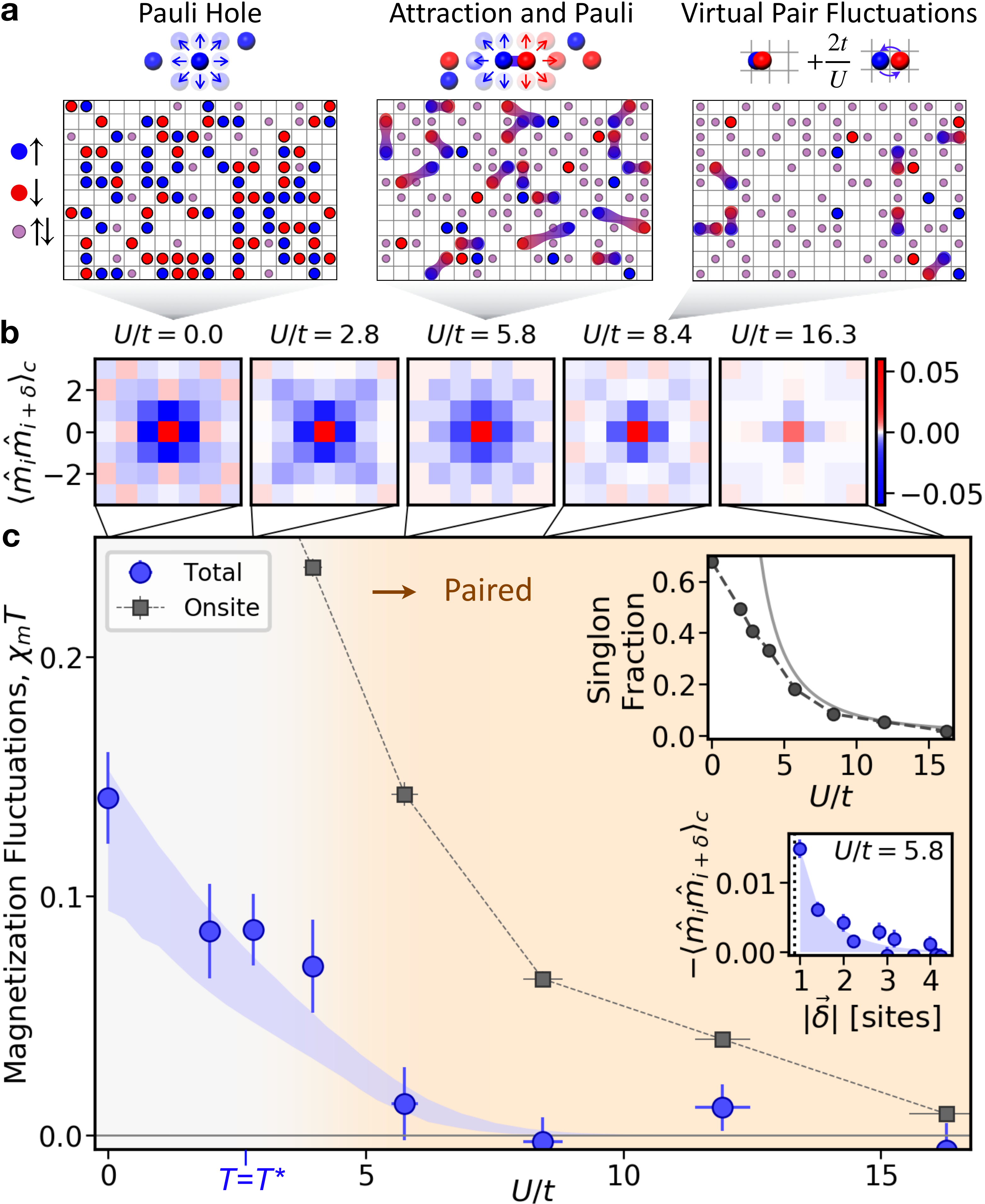}
	\caption{\textbf{Observation of non-local fermion pairing.}
	(a)~Experimental snapshots of the Fermi gas at $U/t{=}0$, $U/t{=}5.8(3)$, and $U/t{=}8.4(4)$~(left to right), showing the formation of non-local pairs and on-site pairs with increasing attraction. Schematics above highlight the physics dominating spin correlations in each image, and shaded bonds suggest possible pair correlations. 
    (b)~Correlation maps $\langle \hat{m}_{i} \hat{m}_{i+\delta}\rangle_c$ of the magnetization $m {=}n_\uparrow {-} n_\downarrow$ at various $U/t$.
	(c)~Total magnetization fluctuations~$\sum_{\vec{\delta}} \langle \hat{m}_{i} \hat{m}_{i+\delta}\rangle_c$~(blue circles) and onsite fluctuations~(black squares) vs.~$U/t$. 
	Total fluctuations equal the product of magnetic susceptibility $\chi_{m}$ and temperature $T$ via the fluctuation-dissipation theorem~\cite{SI}.
	Vanishing total spin fluctuations for $U/t {\gtrsim} 6$~(orange shading) indicate full pairing and vanishing $\chi_{m}$. 
	Blue shading shows quantum Monte Carlo simulations of total fluctuations at $n{=}0.85$, from $T/t{=}0.3$ to $T/t{=}0.4$~\cite{SI}. 
	The pairing temperature $T^*$ crosses $T{\approx} 0.35\, t$ at $U/t{\approx} 2.5$.
	Non-local pairing is reflected in the singlon fraction per total density $s/n$~(upper inset), which scales as ${\sim}8t^2/U^2$~(gray line) at large attraction.
	Fluctuations at $U/t{=}5.8(3)$~(lower inset) extend beyond the interparticle spacing $1/\sqrt{\pi n_\uparrow}$~(dotted line).
	All data and error bars are obtained from bootstrapping greater than 50 images of atomic clouds with imaging loss correction~\cite{SI}.
	}
	\label{fig:Fig_SpinFluctuationSummary}
\end{figure}

\begin{figure*}[!t]
	\centering
	\includegraphics[width=5 in]{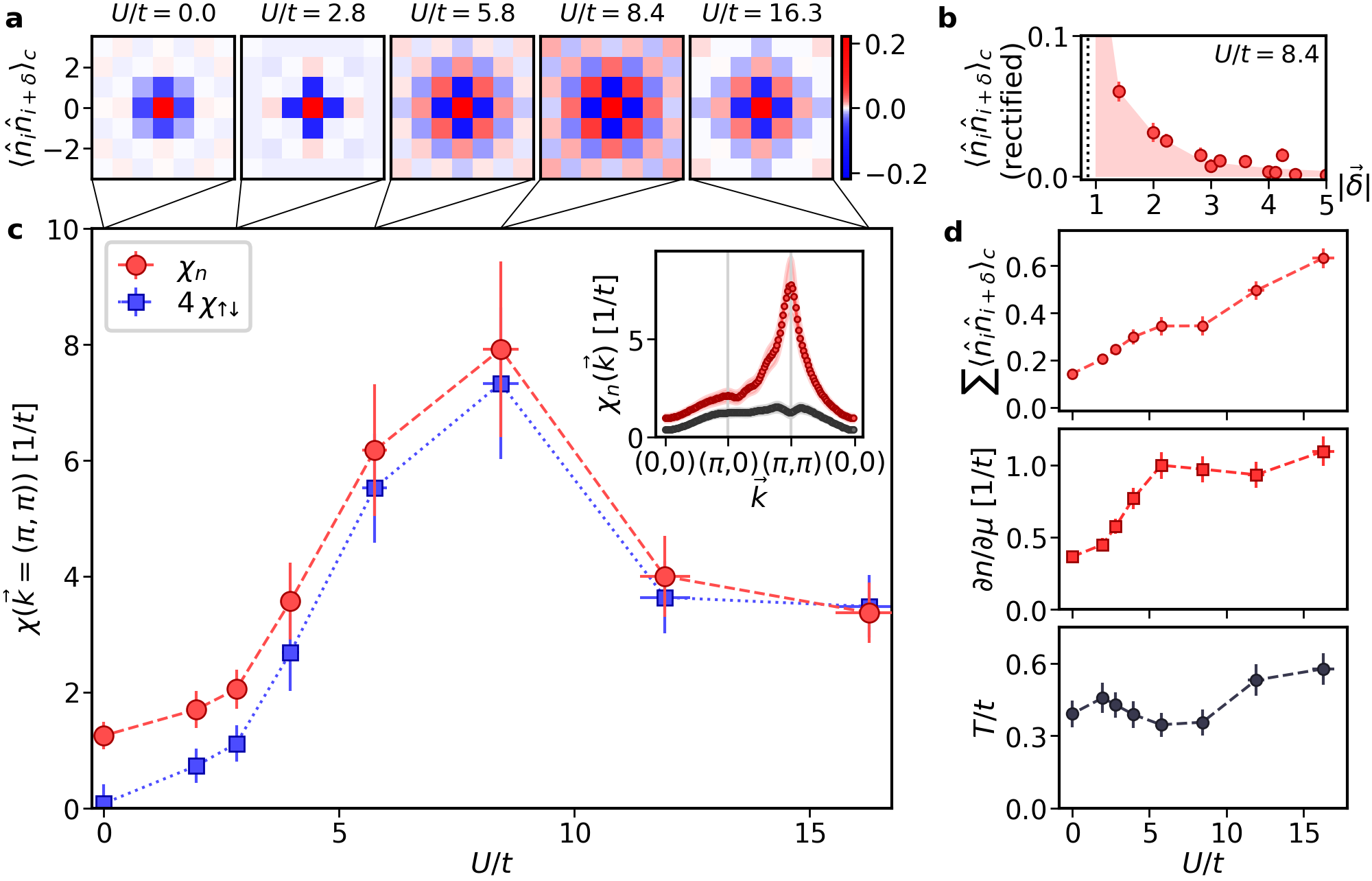}
	\caption{\textbf{Charge-density-wave ordering of pairs.} 
	(a)~Density correlations $\langle \hat{n}_i \hat{n}_{i+\delta}\rangle_c$ reflect the crossover from a non-interacting gas~(left), to a fully paired gas with charge-density-wave order~(center), to a weakly-ordered gas of local pairs~(right).
	(b)~At $U/t{=}8.4(4)$, the non-local rectified correlations $\langle \hat{n}_i \hat{n}_{i+\delta}\rangle_c(-1)^{\delta_x + \delta_y}$ are well described by long-range exponential decay. Dotted line shows the interparticle spacing~$1/\sqrt{\pi n_\uparrow}$.
	(c)~The density response $\chi_{n}$~(red circles) and inter-spin response $\chi_{\uparrow\downarrow}$~(blue squares) at wavevector $\vec{k}{=}(\pi,\pi)$ serve as order parameters for CDW correlations. These susceptibilities are obtained via the fluctuation-dissipation theorem $\chi_{n}(\vec{k}) {=} (1/T) \sum_{\vec{\delta}}\langle \hat{n}_i \hat{n}_{i+\delta}\rangle_c \text{cos} (\vec{k}{\cdot} \vec{\delta})$ and $\chi_{\uparrow \downarrow}(\vec{k}) {=} (1/T) \sum_{\vec{\delta}}\langle \hat{n}_{i\uparrow} \hat{n}_{i+\delta \downarrow}\rangle_c \text{cos} (\vec{k}{\cdot} \vec{\delta})$~\cite{SI}.
	(inset)~$\chi_{n}(\vec{k})$ vs.~$\vec{k}$ at $U/t{=}0$~(black) and $U/t{=}8.4(4)$~(red). 
	(d)~Fluctuation thermometry: The measured density fluctuations $\sum_{\vec{\delta}} \langle \hat{n}_i \hat{n}_{i+\delta}\rangle_c$~(top) and uniform compressibility $\partial n/\partial \mu{=}\chi_{n}(\vec{k}{=}(0,0))$~(middle) are combined to directly obtain the temperature $T/t$~(bottom).
	}
	\label{fig:Fig_ChargeFluctuationsSummary}
\end{figure*}


A true signature of pairing that accounts for these non-local pairs is the vanishing of total spin fluctuations.
Indeed, a system in contact with a surrounding particle bath will generally display fluctuations of the total magnetization $M {=} \sum_i \langle \hat{m}_i\rangle$, where the magnetization $m {=}n_\uparrow {-} n_\downarrow$. However, pair formation suppresses spin fluctuations, as pairs do not contribute to $M$, and thus in a fully paired system the variance $\sigma^2_{M}$ vanishes. This variance is measured locally in our quantum gas microscope through the sum of connected correlations $\sigma^2_{M}/({\rm Area}){=}\sum_{\vec{\delta}} \langle \hat{m}_{i} \hat{m}_{i+\delta}\rangle_c$, where $\langle \hat{m}_{i} \hat{m}_{i+\delta}\rangle_c=\langle \hat{m}_{i} \hat{m}_{i+\delta}\rangle {-}\langle \hat{m}_{i} \rangle \langle \hat{m}_{i+\delta}\rangle$.

The magnetization fluctuations are directly connected to the magnetic susceptibility $\chi_{m} {=} \partial m/\partial h$, the response of the magnetization to a global magnetic field $h$, through the fluctuation-dissipation theorem $\chi_{m} T= \sum_{\vec{\delta}} \langle \hat{m}_{i} \hat{m}_{i+\delta}\rangle_c$~\cite{SI}. 
An energy gap for spin excitations, which exponentially suppresses excess spins and thus $\sum_{\vec{\delta}} \langle \hat{m}_{i} \hat{m}_{i+\delta}\rangle_c$, also exponentially suppresses $\chi_m$~\cite{Yosida1958Paramagnetic}. 

Fig.~\ref{fig:Fig_SpinFluctuationSummary} reports a crossover to full fermion pairing beyond an interaction strength $U/t {\gtrsim} 6$ at $n{=}0.8(1)$ and $T/t{=}0.35(5)$, determined by in-situ observation of magnetization fluctuations. The reduction in fluctuations is in good agreement with theoretical predictions for these parameters~\cite{Singer1996From,Paiva2010Fermions,Fontenele2022Two}.
Fig.~\ref{fig:Fig_SpinFluctuationSummary}(a) highlights the physical mechanisms which determine spin fluctuations at various $U/t$. At vanishing interactions, Pauli exclusion separately reduces the density fluctuations of each spin, and thereby also reduces total spin fluctuations. 
With increasing attraction, non-local pairs form in which spins are subject to a competition of Pauli exclusion and attraction, while deep in the on-site pair regime spin fluctuations reflect virtual hopping onto neighboring sites. From statistical averages over more than 50 spin configurations as in Fig.~\ref{fig:Fig_SpinFluctuationSummary}(a) for each interaction strength, we obtain the two-dimensional magnetization correlation maps $\langle \hat{m}_{i} \hat{m}_{i+\delta}\rangle_c$ shown in Fig.~\ref{fig:Fig_SpinFluctuationSummary}(b). 
To detect pairing, Fig.~\ref{fig:Fig_SpinFluctuationSummary}(c) presents the sum of these correlation maps, the total magnetization fluctuations, which are fully suppressed beyond $U/t {\gtrsim} 6$. 
Already at zero interactions, Pauli exclusion reduces total fluctuations by $68(5)\%$ compared to the high-temperature expectation $n(1{-}n/2)$. This reflects the significant degeneracy of the Fermi gas~($T/T_F {\approx} 0.1$, where $T_F$ is the Fermi temperature).
Increasing attraction reduces magnetization fluctuations further, and the fraction of unpaired spins is less than $1.5(1.8)\%$ at $U/t{=}5.8(3)$, where $\sum_{\vec{\delta}} \langle \hat{m}_{i} \hat{m}_{i+\delta}\rangle_c$ gives the density of unpaired spins. 
This full suppression is dual to the formation of a Mott insulator for repulsive interactions~\cite{Ho2009Quantum, Hartke2020Doublon, Gall2020Simulating, SI}.

The suppression of fluctuations in Fig.~\ref{fig:Fig_SpinFluctuationSummary}(c) with increasing $U/t$ signifies the development of an energy gap for spin excitations~\cite{Yosida1958Paramagnetic}. 
Theory predicts~\cite{Randeria1990Superconductivity,Drechsler1992Crossover,Singer1996From,Singer1998On, Paiva2004Critical,Paiva2010Fermions,Fontenele2022Two} a pairing temperature $T^* {\approx} 0.25\, U$ in the crossover regime~\cite{SI}. This predicted $T^*$ crosses $T{\approx}0.35\,t$ near $U/t{\approx}2.5$, explaining the near complete suppression of fluctuations for $U/t{\gtrsim} 6$. The corresponding expected spin excitation gap far exceeds the two-body binding energy $E_b$, highlighting the many body nature of pairing.

Within this regime of full pairing, a metric for the non-locality of pairs is the singlon density $s$.
The non-local fraction of a pair is $s/n$~(shown in Fig.~\ref{fig:Fig_SpinFluctuationSummary}(c),~upper~inset). The observed scaling of $s/n$ with $t^2/U^2$ at strong attraction is expected from perturbation theory already for a Fermi-Hubbard double well~\cite{Murmann2015Two, Hartke2020Doublon}. At $U/t{=}5.8(3)$, the non-local portion of the pairs amounts to approximately $20\%$. 
The effective size of fermion pairs can be obtained as the spatial extent of non-local spin fluctuations. With full pairing at $U/t {=} 5.8(3)$, spin fluctuations are present beyond the single-spin interparticle spacing~(Fig.~\ref{fig:Fig_SpinFluctuationSummary}(c),~lower~inset), indicating that fermion pairs overlap significantly.


Characteristic for the pseudo-gap regime is a predicted strong departure from Fermi liquid behavior, in which spin and charge fluctuations are similar~\cite{Randeria1992Pairing,Trivedi1995Deviations}.
Having established the existence of non-local fermion pairs through vanishing magnetization fluctuation, we therefore now explore charge (i.e.~density) correlations of the gas. While for weak interactions charge and spin correlations go hand in hand, for stronger attraction we instead find spatial ordering into a charge-density-wave across the phase diagram of Fig.~\ref{fig:Fig_AttractivePhaseDiagram}(a). 
Previously, evidence for charge-density-wave order has only been observed in doublon-doublon correlations and at a fixed interaction strength~\cite{Mitra2018Quantum}.
In Fig.~\ref{fig:Fig_ChargeFluctuationsSummary}(a), beginning without interactions, we observe negative non-local density correlations $\langle \hat{n}_i \hat{n}_{i+\delta}\rangle_c$ for nearest-neighbor and diagonal correlations. These correlations directly equal twice the Pauli hole of a single spin~\cite{Cheuk2016ObservationSpatial,Hartke2020Doublon}, as the measured inter-spin correlations  $\langle \hat{n}_{i \uparrow} \hat{n}_{j \downarrow} \rangle_c$ vanish~\cite{SI}. For increasing attraction, the Pauli hole gives way to the positive checkerboard long-range density correlations, shown vs.~distance in Fig.~\ref{fig:Fig_ChargeFluctuationsSummary}(b) at $U/t{=}8.4(4)$. 
Further increase in $U/t$ reduces the observed CDW strength, likely as a result of smaller effective repulsion between more-localized pairs~(Fig.~\ref{fig:Fig_SpinFluctuationSummary}(b)).
Importantly, we measure a negative sum of non-local inter-spin correlations $\langle \hat{n}_{i \uparrow}\hat{n}_{i + \delta \downarrow}\rangle_c$ for any attractive interaction~\cite{SI}, revealing that a single $\uparrow$ atom in total repels spin $\downarrow$ atoms on all other sites.
This constitutes a strong direct signature of effective repulsion between pairs. 

\begin{figure}[!t]
	\centering
	\includegraphics[width=\columnwidth]{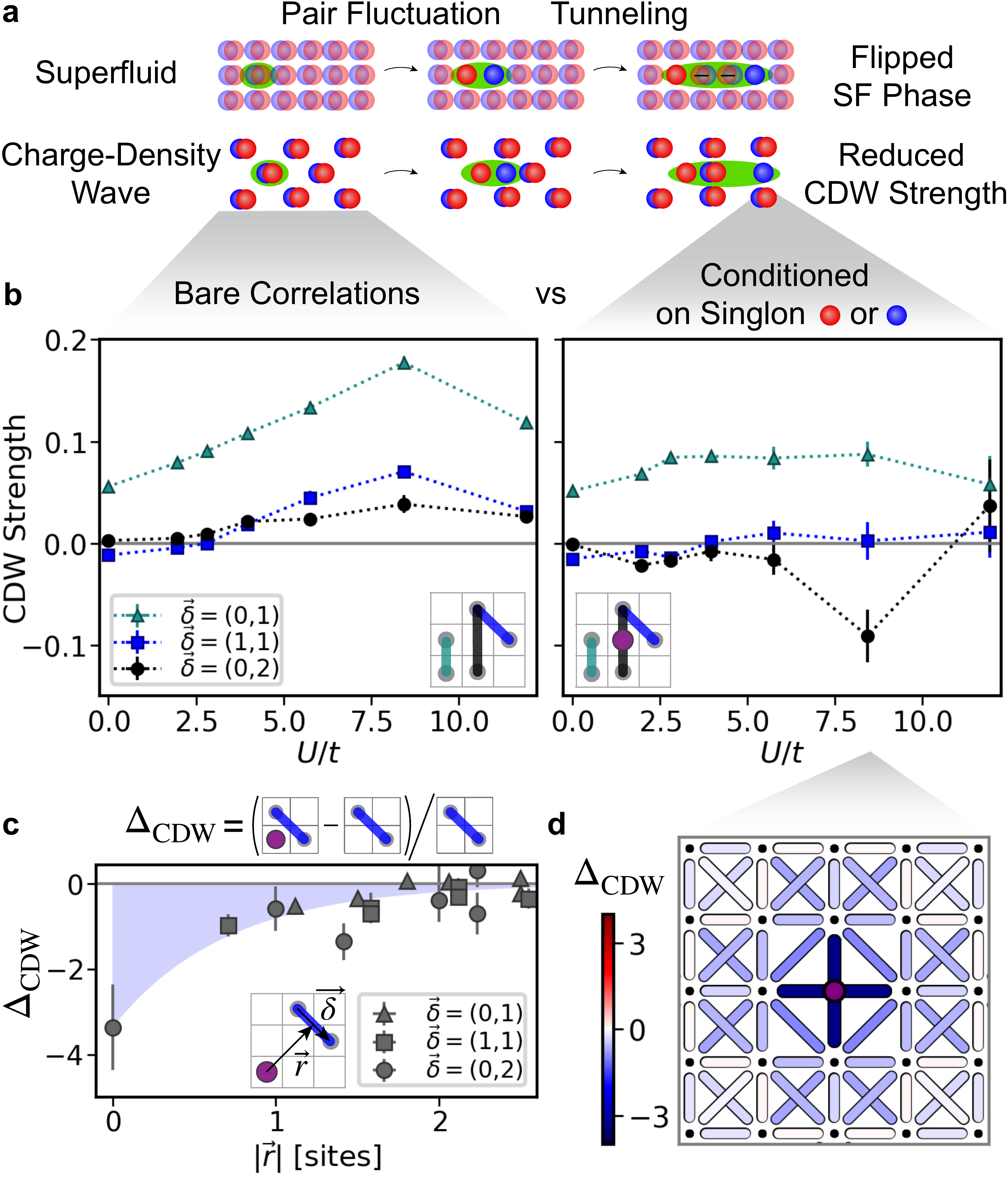}
	\caption{\textbf{Interplay of non-local pairs and many-body order.} 
	(a)~When local pairs in a strongly-correlated attractive system virtually fluctuate into spatially separated single spins, further tunnelling can disturb the underlying many-body ordered state.
	(b)~Polaronic correlations are observed by measuring disturbances of the CDW strength, defined as $\langle (\hat{d}_i{-}\hat{h}_i)(\hat{d}_{i + \delta}{-}\hat{h}_{i + \delta})\rangle_c(-1)^{\delta_x + \delta_y}$, in the vicinity of a single spin. 
	While a peak in the background CDW strength is observed near $U/t{\approx}8$ for various displacements $\vec{\delta}$~(b,~left side), a strong reduction in CDW strength is observed after conditioning on the presence of a nearby isolated spin~(b,~right side). Right inset shows the relative location of the single spin~(purple circle) and CDW bond.
	(c)~The relative change in CDW strength~$\Delta_{\rm CDW}$ reveals the spatial extent of polaronic correlations at $U/t{=}8.4(4)$.  
	$\Delta_{\rm CDW}$ decays with distance $|\vec{r}|$ from the singlon to the CDW bond. 
	The shaded region represents the extent of the disturbance to the background order.
	(d)~A two-dimensional map of (c), with lattice sites represented by black dots and the isolated spin at center. 
	In the CDW strength, $\hat{d}_i$~($\hat{h}_i$) denotes a doublon~(hole).
	}
	\label{fig:Fig_Polaron}
\end{figure}

The development and destruction of CDW order across the phase diagram of Fig.~\ref{fig:Fig_AttractivePhaseDiagram} can be captured by the density response $\chi_{n}$ at wavevector $\vec{k}{=} (\pi,\pi)$~(Fig.~\ref{fig:Fig_ChargeFluctuationsSummary}(b)), which reflects the prevalence of low-energy states with checkerboard order.
Highlighting the power of quantum gas microscopy, this thermodynamic property can be measured in equilibrium using the fluctuation-dissipation theorem for density perturbations, $\chi_{n}(\vec{k}) {=} (1/T) \sum_{\vec{\delta}}\langle \hat{n}_i \hat{n}_{i+\delta}\rangle_c {\text{cos} (\vec{k}{\cdot} \vec{\delta})}$, and the measured uniform density compressibility $\chi_{n}(\vec{k}{=}(0,0)){=}\partial n/\partial \mu$~(Fig.~\ref{fig:Fig_ChargeFluctuationsSummary}(d))~\cite{Hartke2020Doublon}. 
This same method allows measurement of $\chi_{n}(\vec{k})$ throughout the Brillouin zone~(Fig.~\ref{fig:Fig_ChargeFluctuationsSummary}(c)~inset) and provides a model-independent measurement of temperature $T$~(Fig.~\ref{fig:Fig_ChargeFluctuationsSummary}(d))~\cite{Hartke2020Doublon,Zhou2011Universal}. The latter enables us to obtain the magnetic susceptibility $\chi_m$ from the measured spin fluctuations without applying a magnetic field~\cite{Sanner2011Speckle,SI}.
As expected from the phase diagram in Fig.~\ref{fig:Fig_AttractivePhaseDiagram}, the peak in CDW order occurs near $U/t{\approx}6$.
Also displayed are the inter-spin correlations, obtained from  $\chi_{\uparrow \downarrow}(\vec{k}) {=} (1/T) \sum_{\vec{\delta}}\langle \hat{n}_{i\uparrow} \hat{n}_{i+\delta \downarrow}\rangle_c \text{cos} (\vec{k}{\cdot} \vec{\delta})$. While opposite spins are uncorrelated at $U/t{=}0$, they are seen to almost fully carry the CDW order beyond $U/t{\approx} 6$. Since density, magnetization, and inter-spin correlations are related by $\langle \hat{n}_i \hat{n}_{i+\delta}\rangle_c -\langle \hat{m}_i \hat{m}_{i+\delta}\rangle_c = 4\langle \hat{n}_{i \uparrow} \hat{n}_{i+\delta \downarrow}\rangle_c$, the relative agreement of $\chi_{n}$ and $\chi_{\uparrow \downarrow}$ illustrates the strength of density order as compared to magnetic order at $\vec{k}{=} (\pi,\pi)$.
The pronounced CDW peak is also a signature of strong superfluid correlations within the crossover regime, as CDW correlations away from half filling serve as a lower bound for superfluid correlations~\cite{Mitra2018Quantum, Brown2017Spin}.


Given simultaneous charge and spin measurements, we finally explore the interplay of non-local pair fluctuations and the charge-density-wave order of other pairs, revealing the existence of polaronic correlations in the CDW order of the attractive Hubbard model. 
Polaronic correlations occur in the regime of highly non-local pairs, where further tunneling of a separated pair can dislocate the charge-density-wave checkerboard or flip the sign of superfluid correlations~(Fig.~\ref{fig:Fig_Polaron}(a)). These tunneling events prevent the virtual delocalization of other pairs across the bonds where the order has been reversed, costing an additional $4t^2/U$ per bond in the strong-coupling limit, and further confining spatially separated pairs~\cite{Grusdt2018Parton}. 
This mechanism is directly complementary~\cite{Ho2009Quantum, Gall2020Simulating, SI} to the magnetic polaron mechanism of the repulsive Fermi-Hubbard model~\cite{Koepsell2019Imaging,Koepsell2021Microscopic}, though here polaronic correlations dress the individual spins of a spatially separated fermion pair, rather than excess dopants. 

In Fig.~\ref{fig:Fig_Polaron}(b) we compare the charge-density-wave correlations surrounding single spins to those present in the background.
We quantify the underlying CDW strength as $\langle (\hat{d}_i{-}\hat{h}_i)(\hat{d}_{i + \delta}{-}\hat{h}_{i + \delta})\rangle_c(-1)^{\delta_x + \delta_y}$ for a given displacement $\vec{\delta}$, which is positive for any $\vec{\delta}$ for a gas possessing checkerboard doublon-hole correlations. This underlying CDW strength peaks near an interaction strength $U/t{\approx}8$. In contrast, for various $U/t$, the measured CDW strength is strongly reduced after conditioning on the presence of a single nearby isolated spin. This reduction significantly exceeds the lowest order expectation of single pair fluctuation events depicted in Fig.~\ref{fig:Fig_Polaron}(a), e.g.~$25\%$ for a displacement $\vec{\delta}{=}(0,1)$ and
$50\%$ for $\vec{\delta}{=}(1,1)$~or~$(2,0)$.
The measurements directly reveal the spatial extent of these polaronic effects, captured by the relative change $\Delta_{\rm CDW}(|\vec{r}|)$ of conditioned to unconditioned CDW strength, shown in Fig.~\ref{fig:Fig_Polaron}(c-d).
Virtual pair fluctuations disturb the charge-density-wave order over a range of ${\sim}2$~sites at $U/t {=}8.4(4)$, with complete reduction or even reversal of the CDW order on nearby bonds.
In future work, measurements of four-point correlations~\cite{Koepsell2021Microscopic} around pairs of spins will further elucidate the internal structure of these quantum fluctuations.

Our real-space observation of non-local fermion pairing and its interplay with charge-density-wave order illustrates the richness of the pseudo-gap regime of the attractive Hubbard model. Similar competing or intertwined orders are predicted for the repulsive Hubbard model. The methods can be extended further to study polaronic physics and superfluidity~\cite{Chin2006Evidence}, pairing in momentum space as measured in bulk 2D gases~\cite{Holten2022Observation}, to detect the $\pi$ phase shift of CDW order across stripes~\cite{Ying2022Pi}, and to directly measure the BCS condensate fraction through pair correlations~\cite{Ketterle2008MakingFermi}.

{\bf Acknowledgements:}
This work was supported by the NSF through the Center for Ultracold Atoms and Grant PHY-2012110, AFOSR (Grant No. FA9550-16-1-0324 and MURI on molecules No. 2GG016303 PO No15323), and the Vannevar Bush Faculty Fellowship (ONR No. N00014-19-1-2631).


Correspondence and requests for materials should be addressed to T.H.~(hartke@mit.edu) and M.Z.~(zwierlein@mit.edu).

\bibliography{main}

\begin{thebibliography}{48}%
\makeatletter
\providecommand \@ifxundefined [1]{%
 \@ifx{#1\undefined}
}%
\providecommand \@ifnum [1]{%
 \ifnum #1\expandafter \@firstoftwo
 \else \expandafter \@secondoftwo
 \fi
}%
\providecommand \@ifx [1]{%
 \ifx #1\expandafter \@firstoftwo
 \else \expandafter \@secondoftwo
 \fi
}%
\providecommand \natexlab [1]{#1}%
\providecommand \enquote  [1]{``#1''}%
\providecommand \bibnamefont  [1]{#1}%
\providecommand \bibfnamefont [1]{#1}%
\providecommand \citenamefont [1]{#1}%
\providecommand \href@noop [0]{\@secondoftwo}%
\providecommand \href [0]{\begingroup \@sanitize@url \@href}%
\providecommand \@href[1]{\@@startlink{#1}\@@href}%
\providecommand \@@href[1]{\endgroup#1\@@endlink}%
\providecommand \@sanitize@url [0]{\catcode `\\12\catcode `\$12\catcode
  `\&12\catcode `\#12\catcode `\^12\catcode `\_12\catcode `\%12\relax}%
\providecommand \@@startlink[1]{}%
\providecommand \@@endlink[0]{}%
\providecommand \url  [0]{\begingroup\@sanitize@url \@url }%
\providecommand \@url [1]{\endgroup\@href {#1}{\urlprefix }}%
\providecommand \urlprefix  [0]{URL }%
\providecommand \Eprint [0]{\href }%
\providecommand \doibase [0]{http://dx.doi.org/}%
\providecommand \selectlanguage [0]{\@gobble}%
\providecommand \bibinfo  [0]{\@secondoftwo}%
\providecommand \bibfield  [0]{\@secondoftwo}%
\providecommand \translation [1]{[#1]}%
\providecommand \BibitemOpen [0]{}%
\providecommand \bibitemStop [0]{}%
\providecommand \bibitemNoStop [0]{.\EOS\space}%
\providecommand \EOS [0]{\spacefactor3000\relax}%
\providecommand \BibitemShut  [1]{\csname bibitem#1\endcsname}%
\let\auto@bib@innerbib\@empty
\bibitem [{\citenamefont {Inguscio}\ \emph {et~al.}(2008)\citenamefont
  {Inguscio}, \citenamefont {Ketterle},\ and\ \citenamefont
  {Salomon}}]{Inguscio2008Ultracold}%
  \BibitemOpen
  \bibfield  {author} {\bibinfo {author} {\bibfnamefont {M.}~\bibnamefont
  {Inguscio}}, \bibinfo {author} {\bibfnamefont {W.}~\bibnamefont {Ketterle}},
  \ and\ \bibinfo {author} {\bibfnamefont {C.}~\bibnamefont {Salomon}},\
  }\href@noop {} {\enquote {\bibinfo {title} {{Ultracold Fermi Gases}},}\ }
  (\bibinfo {year} {2008})\BibitemShut {NoStop}%
\bibitem [{\citenamefont {Zwerger}(2012)}]{Zwerger2012BCS}%
  \BibitemOpen
  \bibinfo {editor} {\bibfnamefont {W.}~\bibnamefont {Zwerger}},\ ed.,\
  \href@noop {} {\emph {\bibinfo {title} {The BCS-BEC Crossover and the Unitary
  Fermi Gas}}}\ (\bibinfo  {publisher} {Springer Berlin Heidelberg},\ \bibinfo
  {year} {2012})\BibitemShut {NoStop}%
\bibitem [{\citenamefont {Randeria}\ and\ \citenamefont
  {Taylor}(2014)}]{Randeria2014Crossover}%
  \BibitemOpen
  \bibfield  {author} {\bibinfo {author} {\bibfnamefont {M.}~\bibnamefont
  {Randeria}}\ and\ \bibinfo {author} {\bibfnamefont {E.}~\bibnamefont
  {Taylor}},\ }\href {\doibase 10.1146/annurev-conmatphys-031113-133829}
  {\bibfield  {journal} {\bibinfo  {journal} {Annu. Rev. Condens. Matter
  Phys.}\ }\textbf {\bibinfo {volume} {5}},\ \bibinfo {pages} {209} (\bibinfo
  {year} {2014})}\BibitemShut {NoStop}%
\bibitem [{\citenamefont {Randeria}\ \emph {et~al.}(1992)\citenamefont
  {Randeria}, \citenamefont {Trivedi}, \citenamefont {Moreo},\ and\
  \citenamefont {Scalettar}}]{Randeria1992Pairing}%
  \BibitemOpen
  \bibfield  {author} {\bibinfo {author} {\bibfnamefont {M.}~\bibnamefont
  {Randeria}}, \bibinfo {author} {\bibfnamefont {N.}~\bibnamefont {Trivedi}},
  \bibinfo {author} {\bibfnamefont {A.}~\bibnamefont {Moreo}}, \ and\ \bibinfo
  {author} {\bibfnamefont {R.~T.}\ \bibnamefont {Scalettar}},\ }\href {\doibase
  10.1103/PhysRevLett.69.2001} {\bibfield  {journal} {\bibinfo  {journal}
  {Phys. Rev. Lett.}\ }\textbf {\bibinfo {volume} {69}},\ \bibinfo {pages}
  {2001} (\bibinfo {year} {1992})}\BibitemShut {NoStop}%
\bibitem [{\citenamefont {Trivedi}\ and\ \citenamefont
  {Randeria}(1995)}]{Trivedi1995Deviations}%
  \BibitemOpen
  \bibfield  {author} {\bibinfo {author} {\bibfnamefont {N.}~\bibnamefont
  {Trivedi}}\ and\ \bibinfo {author} {\bibfnamefont {M.}~\bibnamefont
  {Randeria}},\ }\href {\doibase 10.1103/PhysRevLett.75.312} {\bibfield
  {journal} {\bibinfo  {journal} {Phys. Rev. Lett.}\ }\textbf {\bibinfo
  {volume} {75}},\ \bibinfo {pages} {312} (\bibinfo {year} {1995})}\BibitemShut
  {NoStop}%
\bibitem [{\citenamefont {Lee}\ \emph {et~al.}(2006)\citenamefont {Lee},
  \citenamefont {Nagaosa},\ and\ \citenamefont {Wen}}]{Lee2006Doping}%
  \BibitemOpen
  \bibfield  {author} {\bibinfo {author} {\bibfnamefont {P.~A.}\ \bibnamefont
  {Lee}}, \bibinfo {author} {\bibfnamefont {N.}~\bibnamefont {Nagaosa}}, \ and\
  \bibinfo {author} {\bibfnamefont {X.-G.}\ \bibnamefont {Wen}},\ }\href
  {\doibase 10.1103/RevModPhys.78.17} {\bibfield  {journal} {\bibinfo
  {journal} {Rev. Mod. Phys.}\ }\textbf {\bibinfo {volume} {78}},\ \bibinfo
  {pages} {17} (\bibinfo {year} {2006})}\BibitemShut {NoStop}%
\bibitem [{\citenamefont {Cao}\ \emph {et~al.}(2018)\citenamefont {Cao},
  \citenamefont {Fatemi}, \citenamefont {Fang}, \citenamefont {Watanabe},
  \citenamefont {Taniguchi}, \citenamefont {Kaxiras},\ and\ \citenamefont
  {Jarillo-Herrero}}]{Cao2018Unconventional}%
  \BibitemOpen
  \bibfield  {author} {\bibinfo {author} {\bibfnamefont {Y.}~\bibnamefont
  {Cao}}, \bibinfo {author} {\bibfnamefont {V.}~\bibnamefont {Fatemi}},
  \bibinfo {author} {\bibfnamefont {S.}~\bibnamefont {Fang}}, \bibinfo {author}
  {\bibfnamefont {K.}~\bibnamefont {Watanabe}}, \bibinfo {author}
  {\bibfnamefont {T.}~\bibnamefont {Taniguchi}}, \bibinfo {author}
  {\bibfnamefont {E.}~\bibnamefont {Kaxiras}}, \ and\ \bibinfo {author}
  {\bibfnamefont {P.}~\bibnamefont {Jarillo-Herrero}},\ }\href {\doibase
  10.1038/nature26160} {\bibfield  {journal} {\bibinfo  {journal} {Nature}\
  }\textbf {\bibinfo {volume} {556}},\ \bibinfo {pages} {43} (\bibinfo {year}
  {2018})}\BibitemShut {NoStop}%
\bibitem [{\citenamefont {Chen}\ \emph {et~al.}(2005)\citenamefont {Chen},
  \citenamefont {Stajic}, \citenamefont {Tan},\ and\ \citenamefont
  {Levin}}]{Chen2005BCS}%
  \BibitemOpen
  \bibfield  {author} {\bibinfo {author} {\bibfnamefont {Q.}~\bibnamefont
  {Chen}}, \bibinfo {author} {\bibfnamefont {J.}~\bibnamefont {Stajic}},
  \bibinfo {author} {\bibfnamefont {S.}~\bibnamefont {Tan}}, \ and\ \bibinfo
  {author} {\bibfnamefont {K.}~\bibnamefont {Levin}},\ }\href {\doibase
  10.1016/j.physrep.2005.02.005} {\bibfield  {journal} {\bibinfo  {journal}
  {Phys. Rep.}\ }\textbf {\bibinfo {volume} {412}},\ \bibinfo {pages} {1}
  (\bibinfo {year} {2005})}\BibitemShut {NoStop}%
\bibitem [{\citenamefont {Scalettar}\ \emph {et~al.}(1989)\citenamefont
  {Scalettar}, \citenamefont {Loh}, \citenamefont {Gubernatis}, \citenamefont
  {Moreo}, \citenamefont {White}, \citenamefont {Scalapino}, \citenamefont
  {Sugar},\ and\ \citenamefont {Dagotto}}]{Scalettar1989Phase}%
  \BibitemOpen
  \bibfield  {author} {\bibinfo {author} {\bibfnamefont {R.~T.}\ \bibnamefont
  {Scalettar}}, \bibinfo {author} {\bibfnamefont {E.~Y.}\ \bibnamefont {Loh}},
  \bibinfo {author} {\bibfnamefont {J.~E.}\ \bibnamefont {Gubernatis}},
  \bibinfo {author} {\bibfnamefont {A.}~\bibnamefont {Moreo}}, \bibinfo
  {author} {\bibfnamefont {S.~R.}\ \bibnamefont {White}}, \bibinfo {author}
  {\bibfnamefont {D.~J.}\ \bibnamefont {Scalapino}}, \bibinfo {author}
  {\bibfnamefont {R.~L.}\ \bibnamefont {Sugar}}, \ and\ \bibinfo {author}
  {\bibfnamefont {E.}~\bibnamefont {Dagotto}},\ }\href {\doibase
  10.1103/PhysRevLett.62.1407} {\bibfield  {journal} {\bibinfo  {journal}
  {Phys. Rev. Lett.}\ }\textbf {\bibinfo {volume} {62}},\ \bibinfo {pages}
  {1407} (\bibinfo {year} {1989})}\BibitemShut {NoStop}%
\bibitem [{\citenamefont {Singer}\ \emph {et~al.}(1998)\citenamefont {Singer},
  \citenamefont {Schneider},\ and\ \citenamefont {Pedersen}}]{Singer1998On}%
  \BibitemOpen
  \bibfield  {author} {\bibinfo {author} {\bibfnamefont {J.}~\bibnamefont
  {Singer}}, \bibinfo {author} {\bibfnamefont {T.}~\bibnamefont {Schneider}}, \
  and\ \bibinfo {author} {\bibfnamefont {M.}~\bibnamefont {Pedersen}},\ }\href
  {\doibase 10.1007/s100510050221} {\bibfield  {journal} {\bibinfo  {journal}
  {Eur. Phys. J. B}\ }\textbf {\bibinfo {volume} {2}},\ \bibinfo {pages} {17}
  (\bibinfo {year} {1998})}\BibitemShut {NoStop}%
\bibitem [{\citenamefont {Paiva}\ \emph {et~al.}(2004)\citenamefont {Paiva},
  \citenamefont {dos Santos}, \citenamefont {Scalettar},\ and\ \citenamefont
  {Denteneer}}]{Paiva2004Critical}%
  \BibitemOpen
  \bibfield  {author} {\bibinfo {author} {\bibfnamefont {T.}~\bibnamefont
  {Paiva}}, \bibinfo {author} {\bibfnamefont {R.~R.}\ \bibnamefont {dos
  Santos}}, \bibinfo {author} {\bibfnamefont {R.~T.}\ \bibnamefont
  {Scalettar}}, \ and\ \bibinfo {author} {\bibfnamefont {P.~J.~H.}\
  \bibnamefont {Denteneer}},\ }\href {\doibase 10.1103/PhysRevB.69.184501}
  {\bibfield  {journal} {\bibinfo  {journal} {Phys. Rev. B}\ }\textbf {\bibinfo
  {volume} {69}},\ \bibinfo {pages} {184501} (\bibinfo {year}
  {2004})}\BibitemShut {NoStop}%
\bibitem [{\citenamefont {Paiva}\ \emph {et~al.}(2010)\citenamefont {Paiva},
  \citenamefont {Scalettar}, \citenamefont {Randeria},\ and\ \citenamefont
  {Trivedi}}]{Paiva2010Fermions}%
  \BibitemOpen
  \bibfield  {author} {\bibinfo {author} {\bibfnamefont {T.}~\bibnamefont
  {Paiva}}, \bibinfo {author} {\bibfnamefont {R.}~\bibnamefont {Scalettar}},
  \bibinfo {author} {\bibfnamefont {M.}~\bibnamefont {Randeria}}, \ and\
  \bibinfo {author} {\bibfnamefont {N.}~\bibnamefont {Trivedi}},\ }\href
  {\doibase 10.1103/PhysRevLett.104.066406} {\bibfield  {journal} {\bibinfo
  {journal} {Phys. Rev. Lett.}\ }\textbf {\bibinfo {volume} {104}},\ \bibinfo
  {pages} {066406} (\bibinfo {year} {2010})}\BibitemShut {NoStop}%
\bibitem [{\citenamefont {Fontenele}\ \emph {et~al.}(2022)\citenamefont
  {Fontenele}, \citenamefont {Costa}, \citenamefont {dos Santos},\ and\
  \citenamefont {Paiva}}]{Fontenele2022Two}%
  \BibitemOpen
  \bibfield  {author} {\bibinfo {author} {\bibfnamefont {R.~A.}\ \bibnamefont
  {Fontenele}}, \bibinfo {author} {\bibfnamefont {N.~C.}\ \bibnamefont
  {Costa}}, \bibinfo {author} {\bibfnamefont {R.~R.}\ \bibnamefont {dos
  Santos}}, \ and\ \bibinfo {author} {\bibfnamefont {T.}~\bibnamefont
  {Paiva}},\ }\href {\doibase 10.1103/PhysRevB.105.184502} {\bibfield
  {journal} {\bibinfo  {journal} {Phys. Rev. B}\ }\textbf {\bibinfo {volume}
  {105}},\ \bibinfo {pages} {184502} (\bibinfo {year} {2022})}\BibitemShut
  {NoStop}%
\bibitem [{\citenamefont {Singer}\ \emph {et~al.}(1996)\citenamefont {Singer},
  \citenamefont {Pedersen}, \citenamefont {Schneider}, \citenamefont {Beck},\
  and\ \citenamefont {Matuttis}}]{Singer1996From}%
  \BibitemOpen
  \bibfield  {author} {\bibinfo {author} {\bibfnamefont {J.~M.}\ \bibnamefont
  {Singer}}, \bibinfo {author} {\bibfnamefont {M.~H.}\ \bibnamefont
  {Pedersen}}, \bibinfo {author} {\bibfnamefont {T.}~\bibnamefont {Schneider}},
  \bibinfo {author} {\bibfnamefont {H.}~\bibnamefont {Beck}}, \ and\ \bibinfo
  {author} {\bibfnamefont {H.-G.}\ \bibnamefont {Matuttis}},\ }\href {\doibase
  10.1103/PhysRevB.54.1286} {\bibfield  {journal} {\bibinfo  {journal} {Phys.
  Rev. B}\ }\textbf {\bibinfo {volume} {54}},\ \bibinfo {pages} {1286}
  (\bibinfo {year} {1996})}\BibitemShut {NoStop}%
\bibitem [{\citenamefont {Moreo}\ and\ \citenamefont
  {Scalapino}(1991)}]{Moreo1991Two}%
  \BibitemOpen
  \bibfield  {author} {\bibinfo {author} {\bibfnamefont {A.}~\bibnamefont
  {Moreo}}\ and\ \bibinfo {author} {\bibfnamefont {D.~J.}\ \bibnamefont
  {Scalapino}},\ }\href {\doibase 10.1103/PhysRevLett.66.946} {\bibfield
  {journal} {\bibinfo  {journal} {Phys. Rev. Lett.}\ }\textbf {\bibinfo
  {volume} {66}},\ \bibinfo {pages} {946} (\bibinfo {year} {1991})}\BibitemShut
  {NoStop}%
\bibitem [{\citenamefont {Keller}\ \emph {et~al.}(1999)\citenamefont {Keller},
  \citenamefont {Metzner},\ and\ \citenamefont
  {Schollw\"ock}}]{Keller1999Thermodynamics}%
  \BibitemOpen
  \bibfield  {author} {\bibinfo {author} {\bibfnamefont {M.}~\bibnamefont
  {Keller}}, \bibinfo {author} {\bibfnamefont {W.}~\bibnamefont {Metzner}}, \
  and\ \bibinfo {author} {\bibfnamefont {U.}~\bibnamefont {Schollw\"ock}},\
  }\href {\doibase 10.1103/PhysRevB.60.3499} {\bibfield  {journal} {\bibinfo
  {journal} {Phys. Rev. B}\ }\textbf {\bibinfo {volume} {60}},\ \bibinfo
  {pages} {3499} (\bibinfo {year} {1999})}\BibitemShut {NoStop}%
\bibitem [{\citenamefont {Bauer}\ \emph {et~al.}(2009)\citenamefont {Bauer},
  \citenamefont {Hewson},\ and\ \citenamefont {Dupuis}}]{Bauer2009Dynamical}%
  \BibitemOpen
  \bibfield  {author} {\bibinfo {author} {\bibfnamefont {J.}~\bibnamefont
  {Bauer}}, \bibinfo {author} {\bibfnamefont {A.~C.}\ \bibnamefont {Hewson}}, \
  and\ \bibinfo {author} {\bibfnamefont {N.}~\bibnamefont {Dupuis}},\ }\href
  {\doibase 10.1103/PhysRevB.79.214518} {\bibfield  {journal} {\bibinfo
  {journal} {Phys. Rev. B}\ }\textbf {\bibinfo {volume} {79}},\ \bibinfo
  {pages} {214518} (\bibinfo {year} {2009})}\BibitemShut {NoStop}%
\bibitem [{\citenamefont {Ho}\ \emph {et~al.}(2009)\citenamefont {Ho},
  \citenamefont {Cazalilla},\ and\ \citenamefont {Giamarchi}}]{Ho2009Quantum}%
  \BibitemOpen
  \bibfield  {author} {\bibinfo {author} {\bibfnamefont {A.~F.}\ \bibnamefont
  {Ho}}, \bibinfo {author} {\bibfnamefont {M.~A.}\ \bibnamefont {Cazalilla}}, \
  and\ \bibinfo {author} {\bibfnamefont {T.}~\bibnamefont {Giamarchi}},\ }\href
  {\doibase 10.1103/PhysRevA.79.033620} {\bibfield  {journal} {\bibinfo
  {journal} {Phys. Rev. A}\ }\textbf {\bibinfo {volume} {79}},\ \bibinfo
  {pages} {033620} (\bibinfo {year} {2009})}\BibitemShut {NoStop}%
\bibitem [{\citenamefont {Gall}\ \emph {et~al.}(2020)\citenamefont {Gall},
  \citenamefont {Chan}, \citenamefont {Wurz},\ and\ \citenamefont
  {K\"ohl}}]{Gall2020Simulating}%
  \BibitemOpen
  \bibfield  {author} {\bibinfo {author} {\bibfnamefont {M.}~\bibnamefont
  {Gall}}, \bibinfo {author} {\bibfnamefont {C.~F.}\ \bibnamefont {Chan}},
  \bibinfo {author} {\bibfnamefont {N.}~\bibnamefont {Wurz}}, \ and\ \bibinfo
  {author} {\bibfnamefont {M.}~\bibnamefont {K\"ohl}},\ }\href {\doibase
  10.1103/PhysRevLett.124.010403} {\bibfield  {journal} {\bibinfo  {journal}
  {Phys. Rev. Lett.}\ }\textbf {\bibinfo {volume} {124}},\ \bibinfo {pages}
  {010403} (\bibinfo {year} {2020})}\BibitemShut {NoStop}%
\bibitem [{\citenamefont {Brown}\ \emph {et~al.}(2020)\citenamefont {Brown},
  \citenamefont {Guardado-Sanchez}, \citenamefont {Spar}, \citenamefont
  {Huang}, \citenamefont {Devereaux},\ and\ \citenamefont
  {Bakr}}]{Brown2020Angle}%
  \BibitemOpen
  \bibfield  {author} {\bibinfo {author} {\bibfnamefont {P.~T.}\ \bibnamefont
  {Brown}}, \bibinfo {author} {\bibfnamefont {E.}~\bibnamefont
  {Guardado-Sanchez}}, \bibinfo {author} {\bibfnamefont {B.~M.}\ \bibnamefont
  {Spar}}, \bibinfo {author} {\bibfnamefont {E.~W.}\ \bibnamefont {Huang}},
  \bibinfo {author} {\bibfnamefont {T.~P.}\ \bibnamefont {Devereaux}}, \ and\
  \bibinfo {author} {\bibfnamefont {W.~S.}\ \bibnamefont {Bakr}},\ }\href
  {\doibase 10.1038/s41567-019-0696-0} {\bibfield  {journal} {\bibinfo
  {journal} {Nat. Phys.}\ }\textbf {\bibinfo {volume} {16}},\ \bibinfo {pages}
  {26} (\bibinfo {year} {2020})}\BibitemShut {NoStop}%
\bibitem [{\citenamefont {Mitra}\ \emph {et~al.}(2018)\citenamefont {Mitra},
  \citenamefont {Brown}, \citenamefont {Guardado-Sanchez}, \citenamefont
  {Kondov}, \citenamefont {Devakul}, \citenamefont {Huse}, \citenamefont
  {Schau{\ss}},\ and\ \citenamefont {Bakr}}]{Mitra2018Quantum}%
  \BibitemOpen
  \bibfield  {author} {\bibinfo {author} {\bibfnamefont {D.}~\bibnamefont
  {Mitra}}, \bibinfo {author} {\bibfnamefont {P.~T.}\ \bibnamefont {Brown}},
  \bibinfo {author} {\bibfnamefont {E.}~\bibnamefont {Guardado-Sanchez}},
  \bibinfo {author} {\bibfnamefont {S.~S.}\ \bibnamefont {Kondov}}, \bibinfo
  {author} {\bibfnamefont {T.}~\bibnamefont {Devakul}}, \bibinfo {author}
  {\bibfnamefont {D.~A.}\ \bibnamefont {Huse}}, \bibinfo {author}
  {\bibfnamefont {P.}~\bibnamefont {Schau{\ss}}}, \ and\ \bibinfo {author}
  {\bibfnamefont {W.~S.}\ \bibnamefont {Bakr}},\ }\href {\doibase
  10.1038/nphys4297} {\bibfield  {journal} {\bibinfo  {journal} {Nat. Phys.}\
  }\textbf {\bibinfo {volume} {14}},\ \bibinfo {pages} {173} (\bibinfo {year}
  {2018})}\BibitemShut {NoStop}%
\bibitem [{\citenamefont {Chan}\ \emph {et~al.}(2020)\citenamefont {Chan},
  \citenamefont {Gall}, \citenamefont {Wurz},\ and\ \citenamefont
  {K{\"{o}}hl}}]{Chan2020Pair}%
  \BibitemOpen
  \bibfield  {author} {\bibinfo {author} {\bibfnamefont {C.~F.}\ \bibnamefont
  {Chan}}, \bibinfo {author} {\bibfnamefont {M.}~\bibnamefont {Gall}}, \bibinfo
  {author} {\bibfnamefont {N.}~\bibnamefont {Wurz}}, \ and\ \bibinfo {author}
  {\bibfnamefont {M.}~\bibnamefont {K{\"{o}}hl}},\ }\href {\doibase
  10.1103/PhysRevResearch.2.023210} {\bibfield  {journal} {\bibinfo  {journal}
  {Phys. Rev. Res.}\ }\textbf {\bibinfo {volume} {2}},\ \bibinfo {pages}
  {023210} (\bibinfo {year} {2020})}\BibitemShut {NoStop}%
\bibitem [{SI()}]{SI}%
  \BibitemOpen
  \href@noop {} {}\bibinfo {note} {See Supplementary Information.}\BibitemShut
  {Stop}%
\bibitem [{\citenamefont {Hartke}\ \emph {et~al.}(2020)\citenamefont {Hartke},
  \citenamefont {Oreg}, \citenamefont {Jia},\ and\ \citenamefont
  {Zwierlein}}]{Hartke2020Doublon}%
  \BibitemOpen
  \bibfield  {author} {\bibinfo {author} {\bibfnamefont {T.}~\bibnamefont
  {Hartke}}, \bibinfo {author} {\bibfnamefont {B.}~\bibnamefont {Oreg}},
  \bibinfo {author} {\bibfnamefont {N.}~\bibnamefont {Jia}}, \ and\ \bibinfo
  {author} {\bibfnamefont {M.}~\bibnamefont {Zwierlein}},\ }\href {\doibase
  10.1103/PhysRevLett.125.113601} {\bibfield  {journal} {\bibinfo  {journal}
  {Phys. Rev. Lett.}\ }\textbf {\bibinfo {volume} {125}},\ \bibinfo {pages}
  {113601} (\bibinfo {year} {2020})}\BibitemShut {NoStop}%
\bibitem [{\citenamefont {Koepsell}\ \emph {et~al.}(2020)\citenamefont
  {Koepsell}, \citenamefont {Hirthe}, \citenamefont {Bourgund}, \citenamefont
  {Sompet}, \citenamefont {Vijayan}, \citenamefont {Salomon}, \citenamefont
  {Gross},\ and\ \citenamefont {Bloch}}]{Koepsell2020Robust}%
  \BibitemOpen
  \bibfield  {author} {\bibinfo {author} {\bibfnamefont {J.}~\bibnamefont
  {Koepsell}}, \bibinfo {author} {\bibfnamefont {S.}~\bibnamefont {Hirthe}},
  \bibinfo {author} {\bibfnamefont {D.}~\bibnamefont {Bourgund}}, \bibinfo
  {author} {\bibfnamefont {P.}~\bibnamefont {Sompet}}, \bibinfo {author}
  {\bibfnamefont {J.}~\bibnamefont {Vijayan}}, \bibinfo {author} {\bibfnamefont
  {G.}~\bibnamefont {Salomon}}, \bibinfo {author} {\bibfnamefont
  {C.}~\bibnamefont {Gross}}, \ and\ \bibinfo {author} {\bibfnamefont
  {I.}~\bibnamefont {Bloch}},\ }\href {\doibase 10.1103/PhysRevLett.125.010403}
  {\bibfield  {journal} {\bibinfo  {journal} {Phys. Rev. Lett.}\ }\textbf
  {\bibinfo {volume} {125}},\ \bibinfo {pages} {010403} (\bibinfo {year}
  {2020})}\BibitemShut {NoStop}%
\bibitem [{\citenamefont {Yosida}(1958)}]{Yosida1958Paramagnetic}%
  \BibitemOpen
  \bibfield  {author} {\bibinfo {author} {\bibfnamefont {K.}~\bibnamefont
  {Yosida}},\ }\href {\doibase 10.1103/PhysRev.110.769} {\bibfield  {journal}
  {\bibinfo  {journal} {Phys. Rev.}\ }\textbf {\bibinfo {volume} {110}},\
  \bibinfo {pages} {769} (\bibinfo {year} {1958})}\BibitemShut {NoStop}%
\bibitem [{\citenamefont {Randeria}\ \emph {et~al.}(1990)\citenamefont
  {Randeria}, \citenamefont {Duan},\ and\ \citenamefont
  {Shieh}}]{Randeria1990Superconductivity}%
  \BibitemOpen
  \bibfield  {author} {\bibinfo {author} {\bibfnamefont {M.}~\bibnamefont
  {Randeria}}, \bibinfo {author} {\bibfnamefont {J.-M.}\ \bibnamefont {Duan}},
  \ and\ \bibinfo {author} {\bibfnamefont {L.-Y.}\ \bibnamefont {Shieh}},\
  }\href {\doibase 10.1103/PhysRevB.41.327} {\bibfield  {journal} {\bibinfo
  {journal} {Phys. Rev. B}\ }\textbf {\bibinfo {volume} {41}},\ \bibinfo
  {pages} {327} (\bibinfo {year} {1990})}\BibitemShut {NoStop}%
\bibitem [{\citenamefont {Drechsler}\ and\ \citenamefont
  {Zwerger}(1992)}]{Drechsler1992Crossover}%
  \BibitemOpen
  \bibfield  {author} {\bibinfo {author} {\bibfnamefont {M.}~\bibnamefont
  {Drechsler}}\ and\ \bibinfo {author} {\bibfnamefont {W.}~\bibnamefont
  {Zwerger}},\ }\href {\doibase 10.1002/andp.19925040105} {\bibfield  {journal}
  {\bibinfo  {journal} {Ann. Phys.}\ }\textbf {\bibinfo {volume} {504}},\
  \bibinfo {pages} {15} (\bibinfo {year} {1992})}\BibitemShut {NoStop}%
\bibitem [{\citenamefont {Murmann}\ \emph {et~al.}(2015)\citenamefont
  {Murmann}, \citenamefont {Bergschneider}, \citenamefont {Klinkhamer},
  \citenamefont {Z\"urn}, \citenamefont {Lompe},\ and\ \citenamefont
  {Jochim}}]{Murmann2015Two}%
  \BibitemOpen
  \bibfield  {author} {\bibinfo {author} {\bibfnamefont {S.}~\bibnamefont
  {Murmann}}, \bibinfo {author} {\bibfnamefont {A.}~\bibnamefont
  {Bergschneider}}, \bibinfo {author} {\bibfnamefont {V.~M.}\ \bibnamefont
  {Klinkhamer}}, \bibinfo {author} {\bibfnamefont {G.}~\bibnamefont {Z\"urn}},
  \bibinfo {author} {\bibfnamefont {T.}~\bibnamefont {Lompe}}, \ and\ \bibinfo
  {author} {\bibfnamefont {S.}~\bibnamefont {Jochim}},\ }\href {\doibase
  10.1103/PhysRevLett.114.080402} {\bibfield  {journal} {\bibinfo  {journal}
  {Phys. Rev. Lett.}\ }\textbf {\bibinfo {volume} {114}},\ \bibinfo {pages}
  {080402} (\bibinfo {year} {2015})}\BibitemShut {NoStop}%
\bibitem [{\citenamefont {Cheuk}\ \emph {et~al.}(2016)\citenamefont {Cheuk},
  \citenamefont {Nichols}, \citenamefont {Lawrence}, \citenamefont {Okan},
  \citenamefont {Zhang}, \citenamefont {Khatami}, \citenamefont {Trivedi},
  \citenamefont {Paiva}, \citenamefont {Rigol},\ and\ \citenamefont
  {Zwierlein}}]{Cheuk2016ObservationSpatial}%
  \BibitemOpen
  \bibfield  {author} {\bibinfo {author} {\bibfnamefont {L.~W.}\ \bibnamefont
  {Cheuk}}, \bibinfo {author} {\bibfnamefont {M.~A.}\ \bibnamefont {Nichols}},
  \bibinfo {author} {\bibfnamefont {K.~R.}\ \bibnamefont {Lawrence}}, \bibinfo
  {author} {\bibfnamefont {M.}~\bibnamefont {Okan}}, \bibinfo {author}
  {\bibfnamefont {H.}~\bibnamefont {Zhang}}, \bibinfo {author} {\bibfnamefont
  {E.}~\bibnamefont {Khatami}}, \bibinfo {author} {\bibfnamefont
  {N.}~\bibnamefont {Trivedi}}, \bibinfo {author} {\bibfnamefont
  {T.}~\bibnamefont {Paiva}}, \bibinfo {author} {\bibfnamefont
  {M.}~\bibnamefont {Rigol}}, \ and\ \bibinfo {author} {\bibfnamefont {M.~W.}\
  \bibnamefont {Zwierlein}},\ }\href {\doibase 10.1126/science.aag3349}
  {\bibfield  {journal} {\bibinfo  {journal} {Science}\ }\textbf {\bibinfo
  {volume} {353}},\ \bibinfo {pages} {1260} (\bibinfo {year}
  {2016})}\BibitemShut {NoStop}%
\bibitem [{\citenamefont {Zhou}\ and\ \citenamefont
  {Ho}(2011)}]{Zhou2011Universal}%
  \BibitemOpen
  \bibfield  {author} {\bibinfo {author} {\bibfnamefont {Q.}~\bibnamefont
  {Zhou}}\ and\ \bibinfo {author} {\bibfnamefont {T.-L.}\ \bibnamefont {Ho}},\
  }\href {\doibase 10.1103/PhysRevLett.106.225301} {\bibfield  {journal}
  {\bibinfo  {journal} {Phys. Rev. Lett.}\ }\textbf {\bibinfo {volume} {106}},\
  \bibinfo {pages} {225301} (\bibinfo {year} {2011})}\BibitemShut {NoStop}%
\bibitem [{\citenamefont {Sanner}\ \emph {et~al.}(2011)\citenamefont {Sanner},
  \citenamefont {Su}, \citenamefont {Keshet}, \citenamefont {Huang},
  \citenamefont {Gillen}, \citenamefont {Gommers},\ and\ \citenamefont
  {Ketterle}}]{Sanner2011Speckle}%
  \BibitemOpen
  \bibfield  {author} {\bibinfo {author} {\bibfnamefont {C.}~\bibnamefont
  {Sanner}}, \bibinfo {author} {\bibfnamefont {E.~J.}\ \bibnamefont {Su}},
  \bibinfo {author} {\bibfnamefont {A.}~\bibnamefont {Keshet}}, \bibinfo
  {author} {\bibfnamefont {W.}~\bibnamefont {Huang}}, \bibinfo {author}
  {\bibfnamefont {J.}~\bibnamefont {Gillen}}, \bibinfo {author} {\bibfnamefont
  {R.}~\bibnamefont {Gommers}}, \ and\ \bibinfo {author} {\bibfnamefont
  {W.}~\bibnamefont {Ketterle}},\ }\href {\doibase
  10.1103/PhysRevLett.106.010402} {\bibfield  {journal} {\bibinfo  {journal}
  {Phys. Rev. Lett.}\ }\textbf {\bibinfo {volume} {106}},\ \bibinfo {pages}
  {010402} (\bibinfo {year} {2011})}\BibitemShut {NoStop}%
\bibitem [{\citenamefont {Brown}\ \emph {et~al.}(2017)\citenamefont {Brown},
  \citenamefont {Mitra}, \citenamefont {Guardado-Sanchez}, \citenamefont
  {Schauß}, \citenamefont {Kondov}, \citenamefont {Khatami}, \citenamefont
  {Paiva}, \citenamefont {Trivedi}, \citenamefont {Huse},\ and\ \citenamefont
  {Bakr}}]{Brown2017Spin}%
  \BibitemOpen
  \bibfield  {author} {\bibinfo {author} {\bibfnamefont {P.~T.}\ \bibnamefont
  {Brown}}, \bibinfo {author} {\bibfnamefont {D.}~\bibnamefont {Mitra}},
  \bibinfo {author} {\bibfnamefont {E.}~\bibnamefont {Guardado-Sanchez}},
  \bibinfo {author} {\bibfnamefont {P.}~\bibnamefont {Schauß}}, \bibinfo
  {author} {\bibfnamefont {S.~S.}\ \bibnamefont {Kondov}}, \bibinfo {author}
  {\bibfnamefont {E.}~\bibnamefont {Khatami}}, \bibinfo {author} {\bibfnamefont
  {T.}~\bibnamefont {Paiva}}, \bibinfo {author} {\bibfnamefont
  {N.}~\bibnamefont {Trivedi}}, \bibinfo {author} {\bibfnamefont {D.~A.}\
  \bibnamefont {Huse}}, \ and\ \bibinfo {author} {\bibfnamefont {W.~S.}\
  \bibnamefont {Bakr}},\ }\href {\doibase 10.1126/science.aam7838} {\bibfield
  {journal} {\bibinfo  {journal} {Science}\ }\textbf {\bibinfo {volume}
  {357}},\ \bibinfo {pages} {1385} (\bibinfo {year} {2017})}\BibitemShut
  {NoStop}%
\bibitem [{\citenamefont {Grusdt}\ \emph {et~al.}(2018)\citenamefont {Grusdt},
  \citenamefont {K\'anasz-Nagy}, \citenamefont {Bohrdt}, \citenamefont {Chiu},
  \citenamefont {Ji}, \citenamefont {Greiner}, \citenamefont {Greif},\ and\
  \citenamefont {Demler}}]{Grusdt2018Parton}%
  \BibitemOpen
  \bibfield  {author} {\bibinfo {author} {\bibfnamefont {F.}~\bibnamefont
  {Grusdt}}, \bibinfo {author} {\bibfnamefont {M.}~\bibnamefont
  {K\'anasz-Nagy}}, \bibinfo {author} {\bibfnamefont {A.}~\bibnamefont
  {Bohrdt}}, \bibinfo {author} {\bibfnamefont {C.~S.}\ \bibnamefont {Chiu}},
  \bibinfo {author} {\bibfnamefont {G.}~\bibnamefont {Ji}}, \bibinfo {author}
  {\bibfnamefont {M.}~\bibnamefont {Greiner}}, \bibinfo {author} {\bibfnamefont
  {D.}~\bibnamefont {Greif}}, \ and\ \bibinfo {author} {\bibfnamefont
  {E.}~\bibnamefont {Demler}},\ }\href {\doibase 10.1103/PhysRevX.8.011046}
  {\bibfield  {journal} {\bibinfo  {journal} {Phys. Rev. X}\ }\textbf {\bibinfo
  {volume} {8}},\ \bibinfo {pages} {011046} (\bibinfo {year}
  {2018})}\BibitemShut {NoStop}%
\bibitem [{\citenamefont {Koepsell}\ \emph {et~al.}(2019)\citenamefont
  {Koepsell}, \citenamefont {Vijayan}, \citenamefont {Sompet}, \citenamefont
  {Grusdt}, \citenamefont {Hilker}, \citenamefont {Demler}, \citenamefont
  {Salomon}, \citenamefont {Bloch},\ and\ \citenamefont
  {Gross}}]{Koepsell2019Imaging}%
  \BibitemOpen
  \bibfield  {author} {\bibinfo {author} {\bibfnamefont {J.}~\bibnamefont
  {Koepsell}}, \bibinfo {author} {\bibfnamefont {J.}~\bibnamefont {Vijayan}},
  \bibinfo {author} {\bibfnamefont {P.}~\bibnamefont {Sompet}}, \bibinfo
  {author} {\bibfnamefont {F.}~\bibnamefont {Grusdt}}, \bibinfo {author}
  {\bibfnamefont {T.~A.}\ \bibnamefont {Hilker}}, \bibinfo {author}
  {\bibfnamefont {E.}~\bibnamefont {Demler}}, \bibinfo {author} {\bibfnamefont
  {G.}~\bibnamefont {Salomon}}, \bibinfo {author} {\bibfnamefont
  {I.}~\bibnamefont {Bloch}}, \ and\ \bibinfo {author} {\bibfnamefont
  {C.}~\bibnamefont {Gross}},\ }\href {\doibase 10.1038/s41586-019-1463-1}
  {\bibfield  {journal} {\bibinfo  {journal} {Nature}\ }\textbf {\bibinfo
  {volume} {572}},\ \bibinfo {pages} {358} (\bibinfo {year}
  {2019})}\BibitemShut {NoStop}%
\bibitem [{\citenamefont {Koepsell}\ \emph {et~al.}(2021)\citenamefont
  {Koepsell}, \citenamefont {Bourgund}, \citenamefont {Sompet}, \citenamefont
  {Hirthe}, \citenamefont {Bohrdt}, \citenamefont {Wang}, \citenamefont
  {Grusdt}, \citenamefont {Demler}, \citenamefont {Salomon}, \citenamefont
  {Gross},\ and\ \citenamefont {Bloch}}]{Koepsell2021Microscopic}%
  \BibitemOpen
  \bibfield  {author} {\bibinfo {author} {\bibfnamefont {J.}~\bibnamefont
  {Koepsell}}, \bibinfo {author} {\bibfnamefont {D.}~\bibnamefont {Bourgund}},
  \bibinfo {author} {\bibfnamefont {P.}~\bibnamefont {Sompet}}, \bibinfo
  {author} {\bibfnamefont {S.}~\bibnamefont {Hirthe}}, \bibinfo {author}
  {\bibfnamefont {A.}~\bibnamefont {Bohrdt}}, \bibinfo {author} {\bibfnamefont
  {Y.}~\bibnamefont {Wang}}, \bibinfo {author} {\bibfnamefont {F.}~\bibnamefont
  {Grusdt}}, \bibinfo {author} {\bibfnamefont {E.}~\bibnamefont {Demler}},
  \bibinfo {author} {\bibfnamefont {G.}~\bibnamefont {Salomon}}, \bibinfo
  {author} {\bibfnamefont {C.}~\bibnamefont {Gross}}, \ and\ \bibinfo {author}
  {\bibfnamefont {I.}~\bibnamefont {Bloch}},\ }\href {\doibase
  10.1126/science.abe7165} {\bibfield  {journal} {\bibinfo  {journal}
  {Science}\ }\textbf {\bibinfo {volume} {374}},\ \bibinfo {pages} {82}
  (\bibinfo {year} {2021})}\BibitemShut {NoStop}%
\bibitem [{\citenamefont {Chin}\ \emph {et~al.}(2006)\citenamefont {Chin},
  \citenamefont {Miller}, \citenamefont {Liu}, \citenamefont {Stan},
  \citenamefont {Setiawan}, \citenamefont {Sanner}, \citenamefont {Xu},\ and\
  \citenamefont {Ketterle}}]{Chin2006Evidence}%
  \BibitemOpen
  \bibfield  {author} {\bibinfo {author} {\bibfnamefont {J.~K.}\ \bibnamefont
  {Chin}}, \bibinfo {author} {\bibfnamefont {D.~E.}\ \bibnamefont {Miller}},
  \bibinfo {author} {\bibfnamefont {Y.}~\bibnamefont {Liu}}, \bibinfo {author}
  {\bibfnamefont {C.}~\bibnamefont {Stan}}, \bibinfo {author} {\bibfnamefont
  {W.}~\bibnamefont {Setiawan}}, \bibinfo {author} {\bibfnamefont
  {C.}~\bibnamefont {Sanner}}, \bibinfo {author} {\bibfnamefont
  {K.}~\bibnamefont {Xu}}, \ and\ \bibinfo {author} {\bibfnamefont
  {W.}~\bibnamefont {Ketterle}},\ }\href {\doibase 10.1038/nature05224}
  {\bibfield  {journal} {\bibinfo  {journal} {Nature}\ }\textbf {\bibinfo
  {volume} {443}},\ \bibinfo {pages} {961} (\bibinfo {year}
  {2006})}\BibitemShut {NoStop}%
\bibitem [{\citenamefont {Holten}\ \emph {et~al.}(2022)\citenamefont {Holten},
  \citenamefont {Bayha}, \citenamefont {Subramanian}, \citenamefont
  {Brandstetter}, \citenamefont {Heintze}, \citenamefont {Lunt}, \citenamefont
  {Preiss},\ and\ \citenamefont {Jochim}}]{Holten2022Observation}%
  \BibitemOpen
  \bibfield  {author} {\bibinfo {author} {\bibfnamefont {M.}~\bibnamefont
  {Holten}}, \bibinfo {author} {\bibfnamefont {L.}~\bibnamefont {Bayha}},
  \bibinfo {author} {\bibfnamefont {K.}~\bibnamefont {Subramanian}}, \bibinfo
  {author} {\bibfnamefont {S.}~\bibnamefont {Brandstetter}}, \bibinfo {author}
  {\bibfnamefont {C.}~\bibnamefont {Heintze}}, \bibinfo {author} {\bibfnamefont
  {P.}~\bibnamefont {Lunt}}, \bibinfo {author} {\bibfnamefont {P.~M.}\
  \bibnamefont {Preiss}}, \ and\ \bibinfo {author} {\bibfnamefont
  {S.}~\bibnamefont {Jochim}},\ }\href {\doibase 10.1038/s41586-022-04678-1}
  {\bibfield  {journal} {\bibinfo  {journal} {Nature}\ }\textbf {\bibinfo
  {volume} {606}},\ \bibinfo {pages} {287} (\bibinfo {year}
  {2022})}\BibitemShut {NoStop}%
\bibitem [{\citenamefont {Ying}\ \emph {et~al.}(2022)\citenamefont {Ying},
  \citenamefont {Scalettar},\ and\ \citenamefont {Mondaini}}]{Ying2022Pi}%
  \BibitemOpen
  \bibfield  {author} {\bibinfo {author} {\bibfnamefont {T.}~\bibnamefont
  {Ying}}, \bibinfo {author} {\bibfnamefont {R.~T.}\ \bibnamefont {Scalettar}},
  \ and\ \bibinfo {author} {\bibfnamefont {R.}~\bibnamefont {Mondaini}},\
  }\href {\doibase 10.1103/PhysRevB.105.115116} {\bibfield  {journal} {\bibinfo
   {journal} {Phys. Rev. B}\ }\textbf {\bibinfo {volume} {105}},\ \bibinfo
  {pages} {115116} (\bibinfo {year} {2022})}\BibitemShut {NoStop}%
\bibitem [{\citenamefont {Ketterle}\ and\ \citenamefont
  {Zwierlein}(2008)}]{Ketterle2008MakingFermi}%
  \BibitemOpen
  \bibfield  {author} {\bibinfo {author} {\bibfnamefont {W.}~\bibnamefont
  {Ketterle}}\ and\ \bibinfo {author} {\bibfnamefont {M.}~\bibnamefont
  {Zwierlein}},\ }\href {\doibase 10.1393/ncr/i2008-10033-1} {\bibfield
  {journal} {\bibinfo  {journal} {Riv. Nuovo Cim.}\ }\textbf {\bibinfo {volume}
  {31}},\ \bibinfo {pages} {247–422} (\bibinfo {year} {2008})}\BibitemShut
  {NoStop}%
\bibitem [{\citenamefont {Cheuk}\ \emph {et~al.}(2015)\citenamefont {Cheuk},
  \citenamefont {Nichols}, \citenamefont {Okan}, \citenamefont {Gersdorf},
  \citenamefont {Ramasesh}, \citenamefont {Bakr}, \citenamefont {Lompe},\ and\
  \citenamefont {Zwierlein}}]{Cheuk2015Quantum}%
  \BibitemOpen
  \bibfield  {author} {\bibinfo {author} {\bibfnamefont {L.~W.}\ \bibnamefont
  {Cheuk}}, \bibinfo {author} {\bibfnamefont {M.~A.}\ \bibnamefont {Nichols}},
  \bibinfo {author} {\bibfnamefont {M.}~\bibnamefont {Okan}}, \bibinfo {author}
  {\bibfnamefont {T.}~\bibnamefont {Gersdorf}}, \bibinfo {author}
  {\bibfnamefont {V.~V.}\ \bibnamefont {Ramasesh}}, \bibinfo {author}
  {\bibfnamefont {W.~S.}\ \bibnamefont {Bakr}}, \bibinfo {author}
  {\bibfnamefont {T.}~\bibnamefont {Lompe}}, \ and\ \bibinfo {author}
  {\bibfnamefont {M.~W.}\ \bibnamefont {Zwierlein}},\ }\href {\doibase
  10.1103/PhysRevLett.114.193001} {\bibfield  {journal} {\bibinfo  {journal}
  {Phys. Rev. Lett.}\ }\textbf {\bibinfo {volume} {114}},\ \bibinfo {pages}
  {193001} (\bibinfo {year} {2015})}\BibitemShut {NoStop}%
\bibitem [{\citenamefont {Hartke}\ \emph {et~al.}(2022)\citenamefont {Hartke},
  \citenamefont {Oreg}, \citenamefont {Jia},\ and\ \citenamefont
  {Zwierlein}}]{Hartke2022Quantum}%
  \BibitemOpen
  \bibfield  {author} {\bibinfo {author} {\bibfnamefont {T.}~\bibnamefont
  {Hartke}}, \bibinfo {author} {\bibfnamefont {B.}~\bibnamefont {Oreg}},
  \bibinfo {author} {\bibfnamefont {N.}~\bibnamefont {Jia}}, \ and\ \bibinfo
  {author} {\bibfnamefont {M.}~\bibnamefont {Zwierlein}},\ }\href {\doibase
  10.1038/s41586-021-04205-8} {\bibfield  {journal} {\bibinfo  {journal}
  {Nature}\ }\textbf {\bibinfo {volume} {601}},\ \bibinfo {pages} {537–541}
  (\bibinfo {year} {2022})}\BibitemShut {NoStop}%
\bibitem [{\citenamefont {Micnas}\ \emph {et~al.}(1990)\citenamefont {Micnas},
  \citenamefont {Ranninger},\ and\ \citenamefont
  {Robaszkiewicz}}]{Micnas1990Superconductivity}%
  \BibitemOpen
  \bibfield  {author} {\bibinfo {author} {\bibfnamefont {R.}~\bibnamefont
  {Micnas}}, \bibinfo {author} {\bibfnamefont {J.}~\bibnamefont {Ranninger}}, \
  and\ \bibinfo {author} {\bibfnamefont {S.}~\bibnamefont {Robaszkiewicz}},\
  }\href {\doibase 10.1103/RevModPhys.62.113} {\bibfield  {journal} {\bibinfo
  {journal} {Rev. Mod. Phys.}\ }\textbf {\bibinfo {volume} {62}},\ \bibinfo
  {pages} {113} (\bibinfo {year} {1990})}\BibitemShut {NoStop}%
\bibitem [{\citenamefont {Bardeen}\ \emph {et~al.}(1957)\citenamefont
  {Bardeen}, \citenamefont {Cooper},\ and\ \citenamefont
  {Schrieffer}}]{Bardeen1957Theory}%
  \BibitemOpen
  \bibfield  {author} {\bibinfo {author} {\bibfnamefont {J.}~\bibnamefont
  {Bardeen}}, \bibinfo {author} {\bibfnamefont {L.~N.}\ \bibnamefont {Cooper}},
  \ and\ \bibinfo {author} {\bibfnamefont {J.~R.}\ \bibnamefont {Schrieffer}},\
  }\href {\doibase 10.1103/PhysRev.108.1175} {\bibfield  {journal} {\bibinfo
  {journal} {Phys. Rev.}\ }\textbf {\bibinfo {volume} {108}},\ \bibinfo {pages}
  {1175} (\bibinfo {year} {1957})}\BibitemShut {NoStop}%
\bibitem [{\citenamefont {Salasnich}\ and\ \citenamefont
  {Toigo}(2012)}]{Salasnich2012Pair}%
  \BibitemOpen
  \bibfield  {author} {\bibinfo {author} {\bibfnamefont {L.}~\bibnamefont
  {Salasnich}}\ and\ \bibinfo {author} {\bibfnamefont {F.}~\bibnamefont
  {Toigo}},\ }\href {\doibase 10.1103/PhysRevA.86.023619} {\bibfield  {journal}
  {\bibinfo  {journal} {Phys. Rev. A}\ }\textbf {\bibinfo {volume} {86}},\
  \bibinfo {pages} {023619} (\bibinfo {year} {2012})}\BibitemShut {NoStop}%
\bibitem [{\citenamefont {Wortis}(1963)}]{Wortis1963Bound}%
  \BibitemOpen
  \bibfield  {author} {\bibinfo {author} {\bibfnamefont {M.}~\bibnamefont
  {Wortis}},\ }\href {\doibase 10.1103/PhysRev.132.85} {\bibfield  {journal}
  {\bibinfo  {journal} {Phys. Rev.}\ }\textbf {\bibinfo {volume} {132}},\
  \bibinfo {pages} {85} (\bibinfo {year} {1963})}\BibitemShut {NoStop}%
\bibitem [{\citenamefont {Miyake}(1983)}]{Miyake1983Fermi}%
  \BibitemOpen
  \bibfield  {author} {\bibinfo {author} {\bibfnamefont {K.}~\bibnamefont
  {Miyake}},\ }\href {\doibase 10.1143/PTP.69.1794} {\bibfield  {journal}
  {\bibinfo  {journal} {Prog. Theor. Phys.}\ }\textbf {\bibinfo {volume}
  {69}},\ \bibinfo {pages} {1794} (\bibinfo {year} {1983})}\BibitemShut
  {NoStop}%
\bibitem [{\citenamefont {Varney}\ \emph {et~al.}(2009)\citenamefont {Varney},
  \citenamefont {Lee}, \citenamefont {Bai}, \citenamefont {Chiesa},
  \citenamefont {Jarrell},\ and\ \citenamefont
  {Scalettar}}]{Varney2009Quantum}%
  \BibitemOpen
  \bibfield  {author} {\bibinfo {author} {\bibfnamefont {C.~N.}\ \bibnamefont
  {Varney}}, \bibinfo {author} {\bibfnamefont {C.-R.}\ \bibnamefont {Lee}},
  \bibinfo {author} {\bibfnamefont {Z.~J.}\ \bibnamefont {Bai}}, \bibinfo
  {author} {\bibfnamefont {S.}~\bibnamefont {Chiesa}}, \bibinfo {author}
  {\bibfnamefont {M.}~\bibnamefont {Jarrell}}, \ and\ \bibinfo {author}
  {\bibfnamefont {R.~T.}\ \bibnamefont {Scalettar}},\ }\href {\doibase
  10.1103/PhysRevB.80.075116} {\bibfield  {journal} {\bibinfo  {journal} {Phys.
  Rev. B}\ }\textbf {\bibinfo {volume} {80}},\ \bibinfo {pages} {075116}
  (\bibinfo {year} {2009})}\BibitemShut {NoStop}%
\end{thebibliography}%

\clearpage

\setcounter{equation}{0}
\setcounter{figure}{0}
\setcounter{secnumdepth}{2}
\renewcommand{\theequation}{S\arabic{equation}}
\renewcommand{\thefigure}{S\arabic{figure}}
\renewcommand{\tocname}{Supplementary Materials}
\renewcommand{\appendixname}{Supplement}



\section*{Supplementary Information}

\subsection{Experimental setup}
The attractive Fermi-Hubbard model is realized from a  degenerate gas comprised of the two lowest hyperfine states of ${}^{40}$K:  $\ket{F{=}9/2, m_F{=}{-}9/2}$ and $\ket{F{=}9/2, m_F{=}{-}7/2}$. 
The atoms occupy a single two-dimensional plane of a three-dimensional optical lattice, with in-plane spacings~$a_x{\approx}a_y{\approx}541\,$nm and out-of-plane spacing $a_z {\approx}3{\,}\mu$m, as described in previous work~\cite{Cheuk2015Quantum, Hartke2020Doublon, Hartke2022Quantum}. 
The amplitude of the in-plane sinusoidal lattice potential is measured to be 4.3(2)$\,E_R$~(recoil energy $E_R=\hbar^2 \pi^2/(2 m a_{x,y}^2) = h{\times} 4260\,$Hz, with $h$ the Planck constant), with a tunneling energy $t{=}h{\times}340(20)\,$Hz.
The out-of-plane $z$ harmonic frequency is $\omega_z {=}2\pi {\times} 4.5(1)\,$kHz.
The envelope of the lattice beams provides an in-plane $x{-}y$ harmonic confinement potential $(1/2)m_{{}^{40}\text{K}}\omega_{x,y}^2 a_{x,y}^2 r^2$, with $r$ the radius in lattice sites and $\omega_{x,y}{=}2\pi {\times} 26.0(9)\,$Hz the mean trapping frequency $\omega_{x,y}^2{=}(\omega_{x}^2 {+} \omega_{y}^2)/2$. 
The atoms interact via $\hat{U}=(4 \pi\hbar^2 a_{\rm 3D}/m) \delta^{(3)}(\textbf{r}_1 {-}\textbf{r}_2)$ with a scattering length $a_{\rm 3D}$ calibrated vs.~magnetic field in Ref.~\cite{Hartke2022Quantum}. We use numerically calculated wavefunctions to obtain $U=\langle \hat{U} \rangle$ when two unlike atoms occupy the same site.

\begin{figure}[!t]
	\centering
	\includegraphics[width=3.5 in]{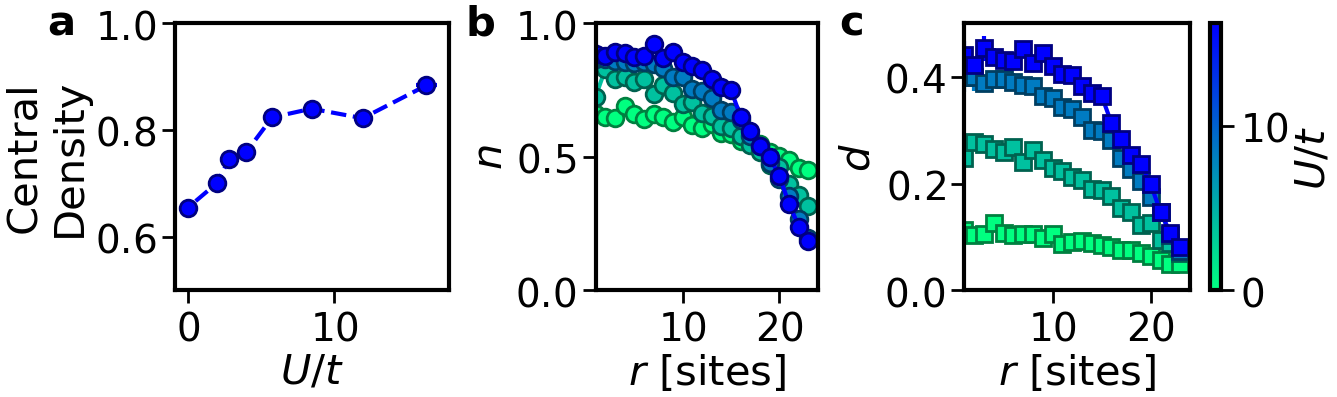}
	\caption{
	(a)~Central density of the atomic cloud vs.~interaction strength $U/t$. 
	Total density~(b) and doublon density~(c) vs.~radius $r$. 
	}
	\label{fig:Fig_DensityVsRadius}
\end{figure}

The central density of the atomic cloud, the density profiles $n(r)$ in the harmonic trap, and the doublon density $d(r)$ are shown in Fig.~\ref{fig:Fig_DensityVsRadius}.  
Throughout the paper, data is obtained from the central region of the cloud of radius 10 sites.

\subsection{A comparison of spin and charge fluctuations}
In a Fermi liquid spin and charge fluctuations go hand in hand. Here we clearly demonstrate diverging behavior of spin and charge with strong pairing in the attractive Hubbard model.

\begin{figure}[!t]
	\centering
	\includegraphics[width=\columnwidth]{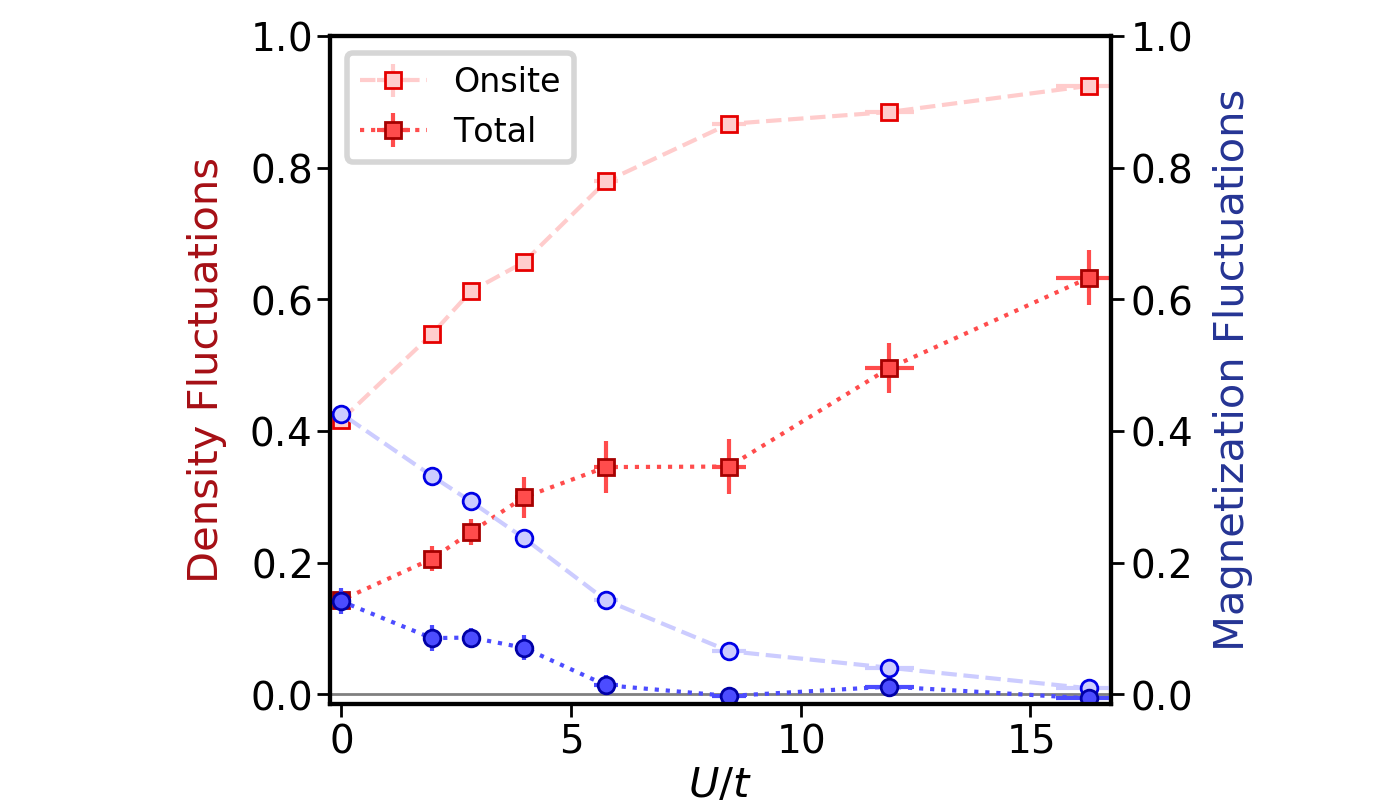}
	\caption{A comparison of density~(red squares) and magnetization~(blue circles) fluctuations from Fig.~\ref{fig:Fig_ChargeFluctuationsSummary}(d) and Fig.~\ref{fig:Fig_SpinFluctuationSummary}(c). 
	Magnetization and density fluctuations are equal at $U/t{=}0$.
	}
	\label{fig:Fig_CombinedFluctuations}
\end{figure}

\begin{figure}[!t]
	\centering
	\includegraphics[width=\columnwidth]{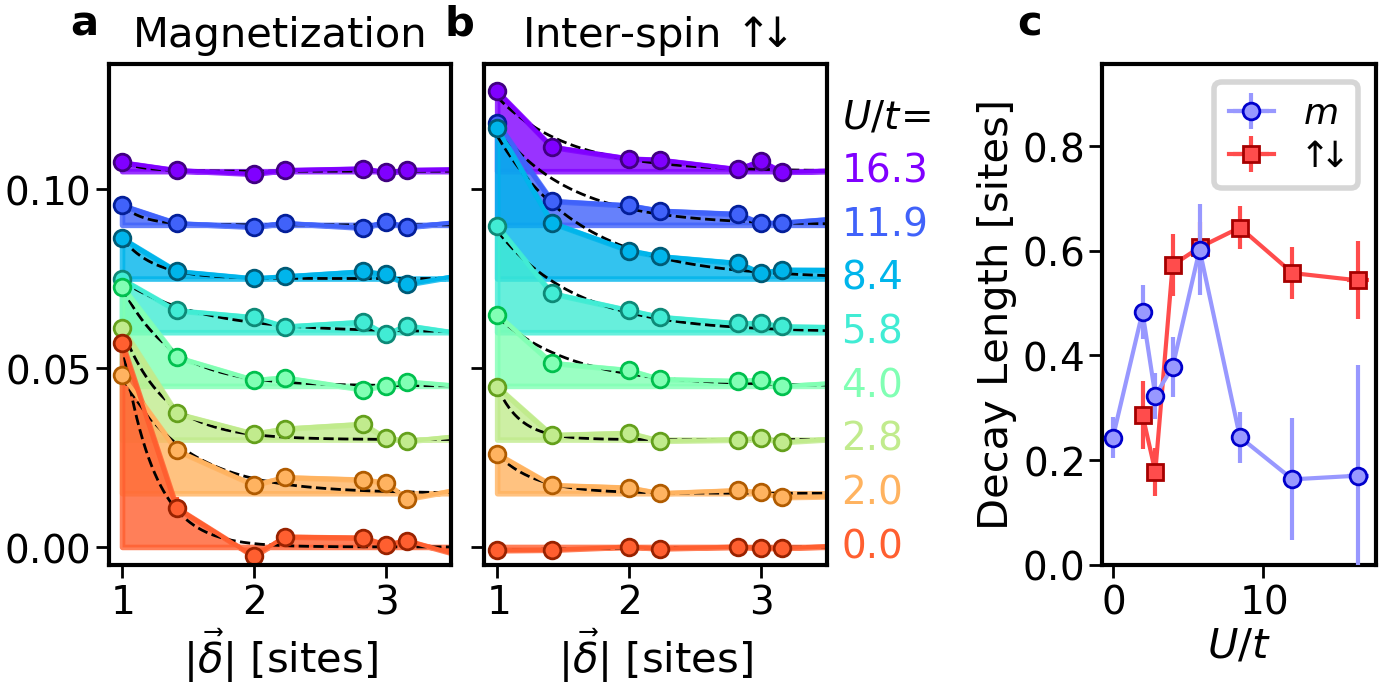}
	\caption{
	(a-b)~Magnetization and inter-spin correlations vs.~displacement $|\vec{\delta}|$ for various $U/t$. 
	(a)~Magnetization correlations $ -\langle \hat{m}_i \hat{m}_{i+\delta}\rangle_c$
	(b)~Rectified inter-spin correlations 
	$\langle \hat{n}_{i \uparrow} \hat{n}_{i+\delta \downarrow} \rangle_c (-1)^{\delta_x + \delta_y}$. 
	(c)~Fitted exponential decay length to the data in (a)~(blue circles) and (b)~(red squares). Data in (a-b) are each offset vertically by 0.015.
	}
	\label{fig:Fig_Waterfall}
\end{figure}

\begin{figure}[!t]
	\centering
	\includegraphics[width=\columnwidth]{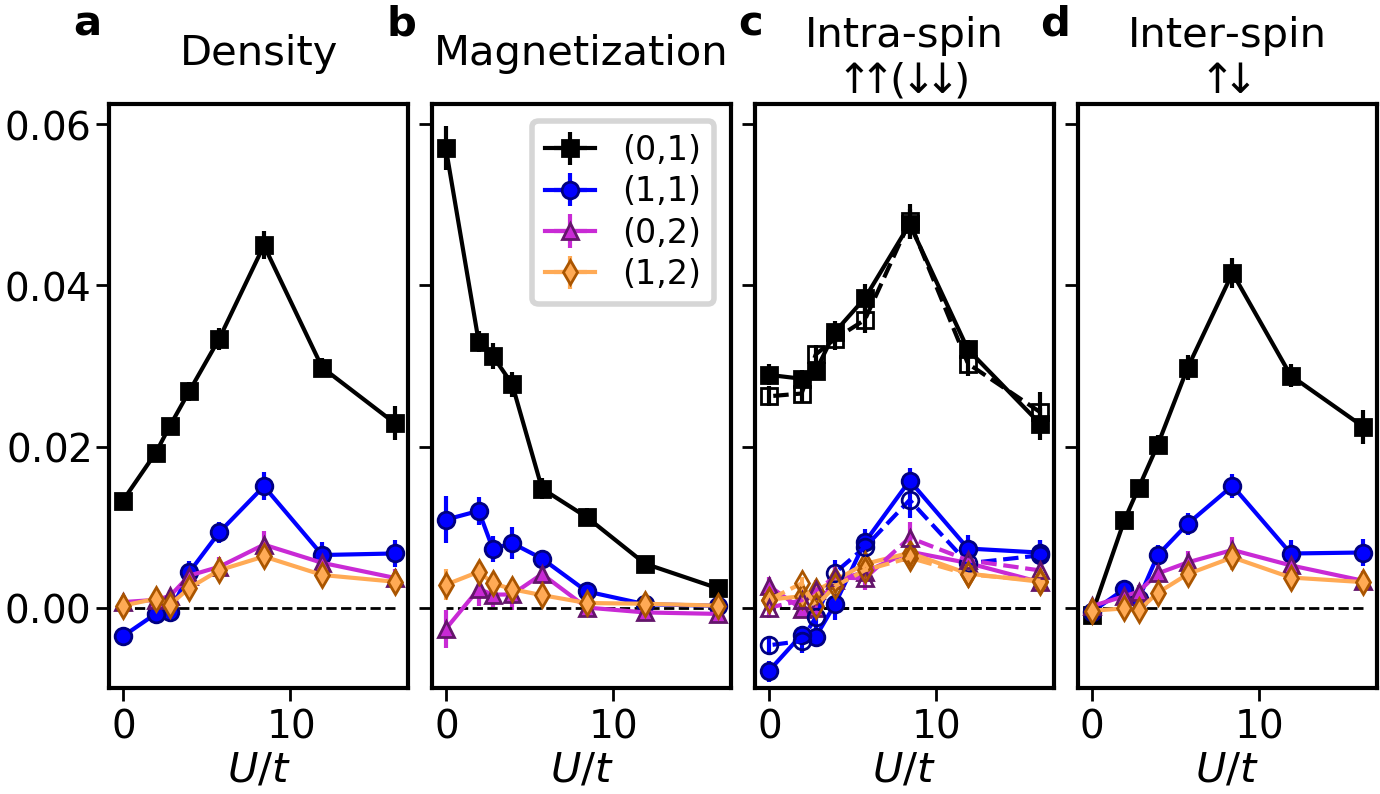}
	\caption{
	(a)~Normalized rectified density correlations $\langle \hat{n}_i \hat{n}_{i+\delta}\rangle_c(-1)^{\delta_x + \delta_y}/4$ for fixed displacements $\vec{\delta}$ vs.~$U/t$. 
	(b)~Magnetization correlations~$-\langle \hat{m}_i \hat{m}_{i+\delta}\rangle_c$. 
	(c)~Rectified intra-spin correlations $\langle \hat{n}_{i \uparrow} \hat{n}_{i+\delta \uparrow} \rangle_c(-1)^{\delta_x + \delta_y}$ (solid) and $\langle \hat{n}_{i \downarrow} \hat{n}_{i+\delta \downarrow} \rangle_c(-1)^{\delta_x + \delta_y}$ (hollow).
	(d)~Rectified inter-spin correlations $\langle \hat{n}_{i \uparrow} \hat{n}_{i+\delta \downarrow} \rangle_c(-1)^{\delta_x + \delta_y}$. 
	}
	\label{fig:Fig_AllCorrelators}
\end{figure}

Fig.~\ref{fig:Fig_CombinedFluctuations} compares the total density and magnetization fluctuations as attractive interactions are increased. At vanishing interactions, the observed equality of total fluctuations of magnetization and density reflects vanishing inter-spin correlations $\langle \hat{n}_{i \uparrow} \hat{n}_{i+\delta \downarrow} \rangle_c$ due to the relation $\langle \hat{n}_i \hat{n}_{i+\delta}\rangle_c -\langle \hat{m}_i \hat{m}_{i+\delta}\rangle_c = 4\langle \hat{n}_{i \uparrow} \hat{n}_{i+\delta \downarrow}\rangle_c$. Vanishing correlations $\langle \hat{n}_{i \uparrow} \hat{n}_{i+\delta \downarrow} \rangle_c$ at $U/t{=}0$ are shown explicitly in Fig.~\ref{fig:Fig_Waterfall}(b) and Fig.~\ref{fig:Fig_AllCorrelators}(d). 
With increasing attraction, spin fluctuations in Fig.~\ref{fig:Fig_CombinedFluctuations} are reduced, and ultimately vanish, while density fluctuations increase, reflecting the remaining fluctuations in spatial organization of paired fermions. 

Magnetization and rectified inter-spin correlations are both well described by an exponential decay vs.~displacement $|\vec{\delta}|$ for all interactions~(Fig.~\ref{fig:Fig_Waterfall}(a-b)), reflecting the existence of finite size pairs and CDW order. 
Inter-spin correlations initially grow with increasing $U/t$, with an amplitude that peaks near $U/t\approx8$~(see Fig.~\ref{fig:Fig_AllCorrelators}(d)), and an exponential decay length that saturates at strong attraction~(Fig.~\ref{fig:Fig_Waterfall}(c)). The magnetization correlations instead are reduced with increasing attraction. The magnetization correlation decay length, which can be interpreted as the pair size, initially grows and peaks near $U/t=5.8(3)$, before sharply decreasing at strong attraction, reflecting the formation of local pairs.

The spin balanced Hubbard model possesses a symmetry between spin $\uparrow$ and spin $\downarrow$, which implies intra-spin correlations $\langle \hat{n}_{i \uparrow} \hat{n}_{i+\delta \uparrow} \rangle_c$ and $\langle \hat{n}_{i \downarrow} \hat{n}_{i+\delta \downarrow } \rangle_c$ should be equal. Fig.~\ref{fig:Fig_AllCorrelators}(c) shows this equality in the measured data for all $U/t$ and various displacements $\vec{\delta}$. 
For reference, Fig.~\ref{fig:Fig_AllCorrelators}(a-b) also show density and magnetization correlations at specific displacements $\vec{\delta}$.

\begin{figure}[t]
	\centering
	\includegraphics[width=\columnwidth]{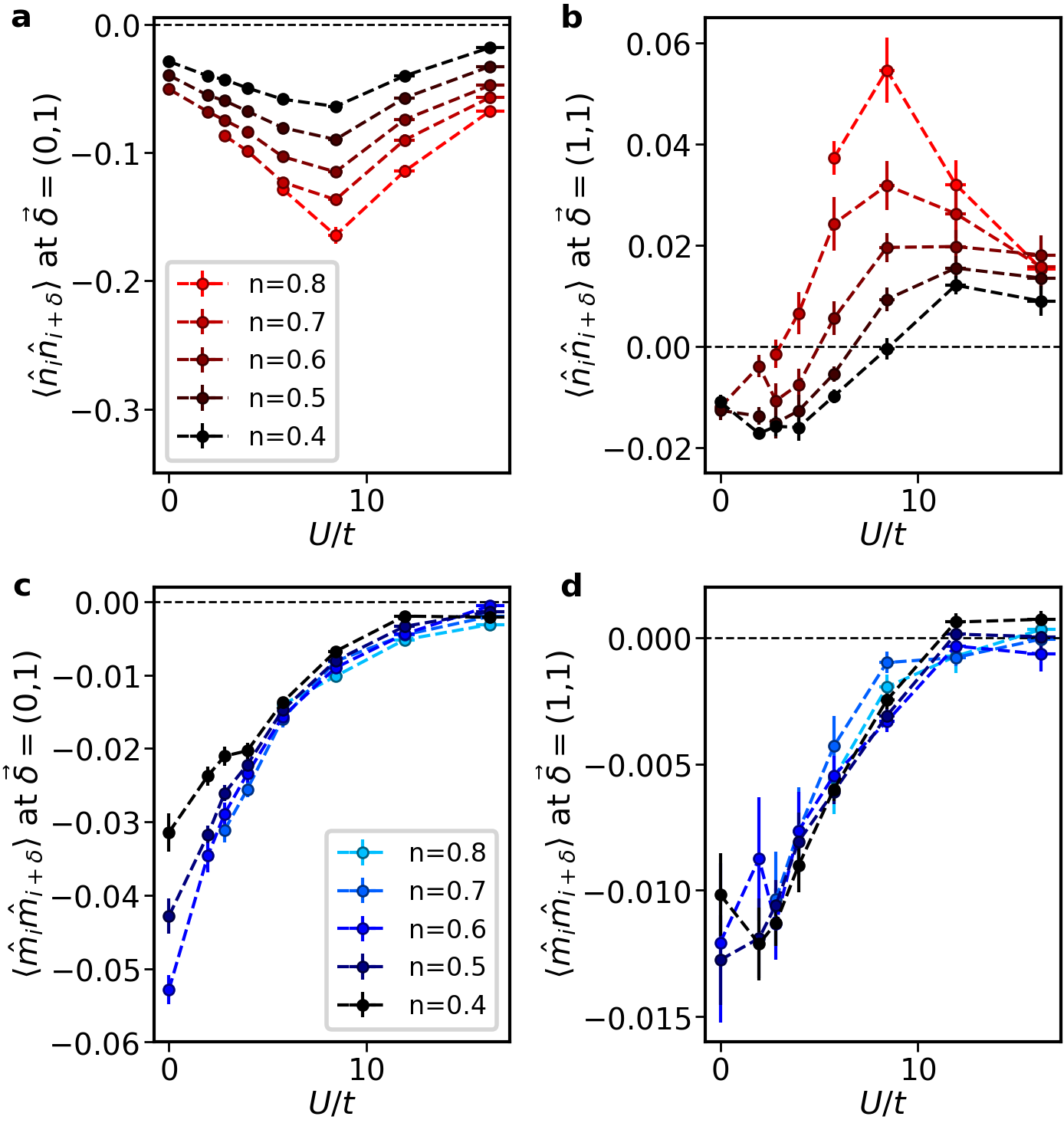}
	\caption{Density dependence of correlations.
	(a-b)~Connected density-density correlations $\langle \hat{n}_i \hat{n}_{i+\delta}\rangle_c$ on the nearest neighbor~(a) and diagonal~(b) at various fixed values of the density $n$. 
	(c-d)~Connected magnetization correlations.
	}
	\label{fig:Fig_CorrelatorsVsDensity}
\end{figure}

The properties of the Fermi-Hubbard model depend significantly on the density of atoms, with a competition of charge-density-wave and superfluid correlations at $n{=}1$, and reduced CDW correlations away from $n=1$~\cite{Mitra2018Quantum}. In Fig.~\ref{fig:Fig_CorrelatorsVsDensity} we show the density dependence of spin and charge correlations for nearest-neighbor and diagonal displacements, obtained from lower density regions of the atomic cloud.
These correlations reveal a competition between CDW order at strong attraction and the Pauli hole at weak attraction through a sign change of diagonal density correlations with increasing $U/t$~(Fig.~\ref{fig:Fig_CorrelatorsVsDensity}(b)). 
At lower densities, the Pauli hole extends over a larger region, and this leads to a persistence of negative diagonal correlations to higher values of $U/t$. These observations are dual to the sign reversal of diagonal magnetic correlations in the spin-imbalanced repulsive Hubbard model at half filling~\cite{Brown2017Spin}.

In contrast, Fig.~\ref{fig:Fig_CorrelatorsVsDensity}(c-d) demonstrate that spin correlations remain largely independent of density, reflecting the origin of spin fluctuations in virtual fluctuations of paired atoms. The suppression of diagonal spin correlations compared to nearest-neighbor correlations reflects the origin of diagonal correlations as a second order tunneling process, suppressed by an additional factor of $t/U$. 

\subsection{The attractive Fermi-Hubbard Hamiltonian}
The Fermi-Hubbard Hamiltonian is 
\begin{multline}
    \hat{H} = -t \sum_{\langle i,j \rangle \sigma}\left( \hat{c}_{i\sigma }^\dagger \hat{c}_{j\sigma } + \text{h.c.} \right)
    - U \sum_{i}\hat{n}_{i \uparrow} \hat{n}_{i \downarrow}
    \\
    - \mu \sum_{i}(\hat{n}_{i \uparrow}+ \hat{n}_{i \downarrow} ) 
    - h \sum_{i}(\hat{n}_{i \uparrow}- \hat{n}_{i \downarrow}),
    \label{eqn:FHGrandCanonicalHamiltonian}
\end{multline}
with tunneling amplitude $t$, attractive onsite interaction $U$, and spin $\sigma \!\! = \, \uparrow$ or $\sigma \!\! =  \,\downarrow$. We work in the grand-canonical description, and include a chemical potential $\mu$ and magnetic field $h$ coupled to the density $n$ and magnetization $m$. 

A mapping exists between attractive and repulsive systems, $\hat{c}_{i_xi_y\downarrow} \leftrightarrow (-1)^{i_x + i_y}\hat{c}^\dagger_{i_x i_y \downarrow}$, which 
leaves the Hamiltonian in Eqn.~(\ref{eqn:FHGrandCanonicalHamiltonian}) unchanged in form, flips the sign of interactions $U \leftrightarrow -U$, and interchanges $\mu \leftrightarrow h {+} U/2$, $h \leftrightarrow \mu {+} U/2$, and  $n-1 \leftrightarrow m$~\cite{Ho2009Quantum, Gall2020Simulating}. This mapping implies that the correlations of isolated spin $\uparrow$ and spin $\downarrow$ atoms in the attractive Hubbard model are equal to the correlations of doublons and holes~(completely empty sites) in the repulsive Hubbard model~\cite{Hartke2020Doublon}, at the appropriately transformed values of $\mu$ and $h$. 

\subsection{Hard core boson limit}

In the limit of strong attraction, the Fermi-Hubbard model is well described by treating pairs as hard core bosons with density $\hat{n}_{b,i}=\hat{b}^\dagger_i \hat{b}_i$ and bosonic creation operators $\hat{b}^\dagger_i$. 
A Fermi-Hubbard double well system can be used to derive the terms of this effective Hamiltonian. 
By combining double well terms for each lattice bond, we obtain, aside from a constant and an effective chemical potential, the Hamiltonian on a lattice~\cite{Micnas1990Superconductivity},
\begin{equation}
    \hat{H}_{\rm b} = -\frac{J}{2} \sum_{\langle i,j \rangle } \left( \hat{b}_{i }^\dagger \hat{b}_j + \text{h.c.} \right)
    +
    J \sum_{\langle i,j \rangle } \hat{n}_{b,i}\hat{n}_{b,j}
    \label{eqn:FHBosonEffectiveModel}
\end{equation}

\begin{figure}[!t]
	\centering
	\includegraphics[width=3.5 in]{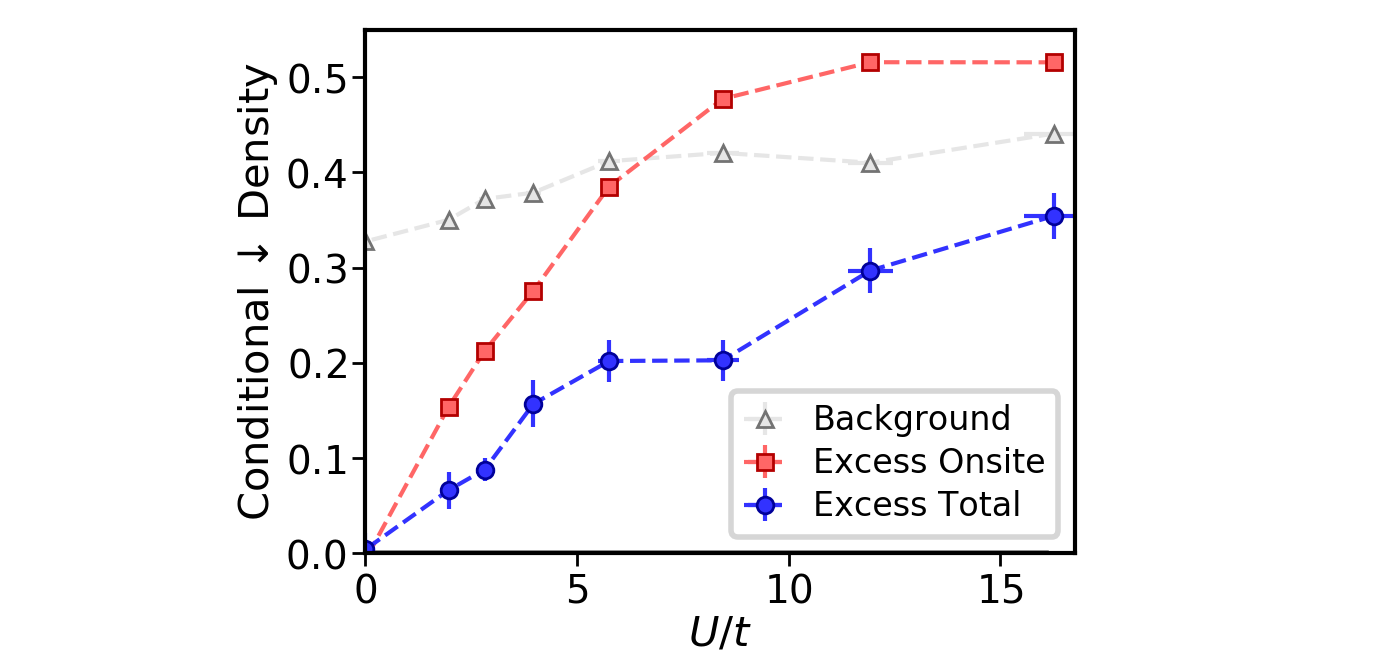}
	\caption{Repulsion between pairs is evident in the dressing cloud of $\downarrow$ spins around a spin $\uparrow$ atom. 
	Spin $\downarrow$ atoms have a background density $n_\downarrow$~(grey triangles).
	Conditioned on the presence of a spin $\uparrow$ atom, there is an excess probability to find a spin $\downarrow$ atom on the same site, $\langle \hat{n}_{i\downarrow} \hat{n}_{i\uparrow} \rangle/ n_{\uparrow} - n_{\downarrow}$~(red squares). Growth of this excess onsite probability with $U/t$ reflects the increasing prevalence of doubly-occupied sites.
	The total conditional excess $\downarrow$ density near a given $\uparrow$ atom $\sum_{\vec{\delta}} (\langle \hat{n}_{i+\delta \downarrow} \hat{n}_{i\uparrow} \rangle/ n_{\uparrow} - n_{\downarrow})$~(blue circles), which includes excess $\downarrow$ atoms onsite and in the surrounding area, is smaller than the onsite excess density. This implies a spin $\uparrow$ atom in total repels opposite spins on all other sites.
	}
	\label{fig:Fig_ExcessDensity}
\end{figure}

This Hamiltonian features nearest neighbor repulsion between pairs. Experimental evidence for this repulsion is shown in Fig.~\ref{fig:Fig_ExcessDensity}, which presents the total excess spin $\downarrow$ found in a local area surrounding a site occupied by a spin $\uparrow$, termed the conditional excess density of spin $\downarrow$. 
The total excess density~(blue circles) is observed to be lower than the onsite excess density~(red squares) for all $U/t$, showing that, although a spin $\uparrow$ attracts a spin $\downarrow$ on the same site, it reduces the total probability of spin $\downarrow$ atoms on nearby sites. 

In addition, measuring the total excess opposite spin surrounding an atom in Fig.~\ref{fig:Fig_ExcessDensity} highlights the intricate nature of pairing in the Hubbard gas. 
As discussed in Fig.~\ref{fig:Fig_SpinFluctuationSummary}(c), full pairing is coincident with vanishing total magnetization fluctuations.
However, Fig.~\ref{fig:Fig_ExcessDensity} shows that each spin attracts only a fraction of nearby opposite spin on net, implying that in order for total magnetization fluctuations to vanish, nearby alike spins must also be repelled due to Pauli exclusion. This follows from the formula for spin balanced magnetization fluctuation sums, 
\begin{equation}
    \sum_{\vec{\delta}} \langle \hat{m}_{i} \hat{m}_{i+\delta}\rangle_c
    = 
    2\sum_{\vec{\delta}}
    \left(\langle \hat{n}_{i\uparrow} \hat{n}_{i+\delta \uparrow} \rangle_c
    -
    \langle \hat{n}_{i\uparrow} \hat{n}_{i+\delta \downarrow} \rangle_c
    \right)
    ,
\end{equation}
where the first term on the right side is affected by Pauli exclusion and the second reflects conditional excess density. This same combination of Pauli exclusion and attraction ensures vanishing magnetization fluctuations in the BCS state. 

\begin{figure}[t]
	\centering
	\includegraphics[width=\columnwidth]{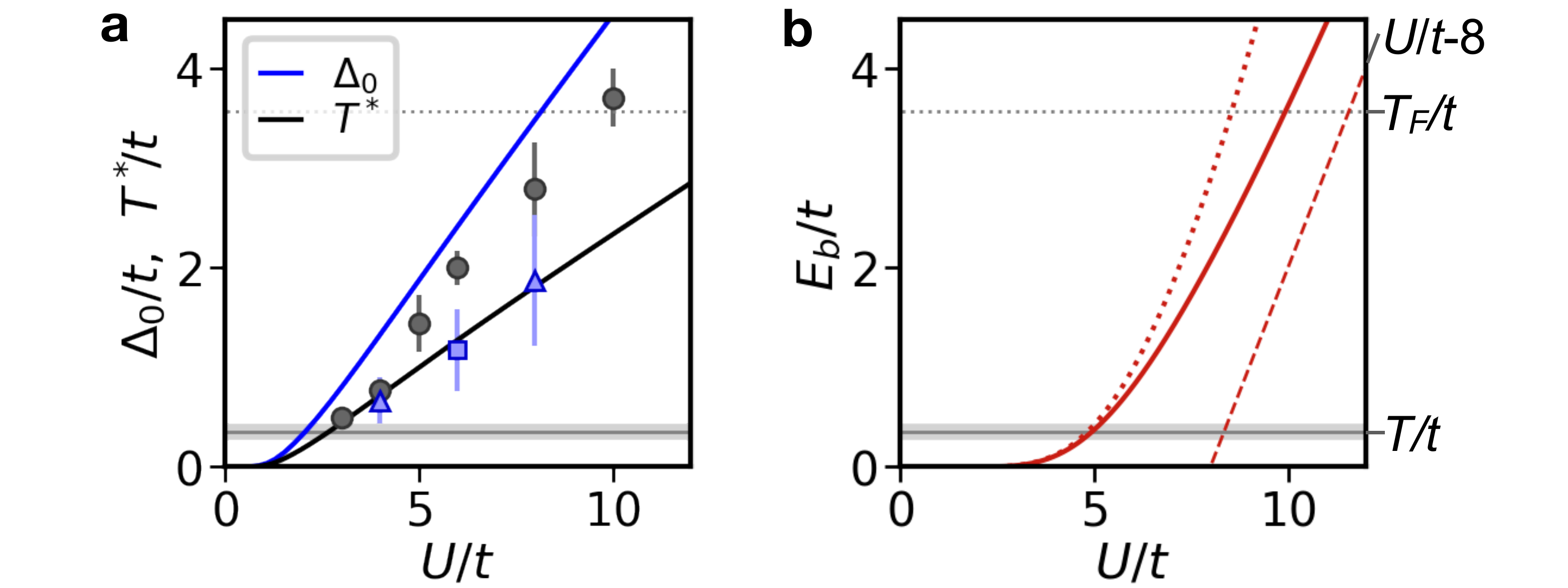}
	\caption{
	(a)~Fermi-Hubbard mean field pairing gap $\Delta_0$~(blue line) and pairing temperature $T^*$~(black line) vs.~$U/t$, calculated at $n{=}0.8$. Also shown are quantum Monte Carlo calculations of $\Delta_0$~(blue squares and triangles, taken from the gap in spectral functions in Refs.~\cite{Singer1996From, Singer1998On}) and the pairing temperature $T_p$ where $\chi_m$ is reduced upon cooling~(grey circles, from Ref.~\cite{Fontenele2022Two} at $n{=}0.87$).
	(b)~Analytic two-body bound state energy $E_b$ in a lattice vs.~$U/t$~(solid line), which approaches $E_b{=}U{-}8t$ at strong attraction~(dashed line), and is approximately $64t\exp(-8\pi t/U)$ at small $U/t$~(dotted line). The experimental temperature $T/t{\approx} 0.35$ and Fermi temperature $T_F/t{\approx} 3.5$ are also marked~(horizontal lines).  
	}
	\label{fig:Fig_BoundState}
\end{figure}

\subsection{Estimates of the pairing temperature $T^*$}
Away from half filling, one approach to gain insight into pairing in the attractive Hubbard model is through a mean-field Ansatz, such as the BCS state~\cite{Bardeen1957Theory, Salasnich2012Pair}. One obtains a pairing gap $\Delta_0$ at zero temperature, and a characteristic temperature $T^*$ for the onset of pairing, given by the temperature where $\Delta(T)$ becomes nonzero upon cooling. Calculated values of $\Delta_0$ and $T^*$ at $n{=}0.8$ are shown in Fig.~\ref{fig:Fig_BoundState}(a), and are compared to quantum Monte Carlo calculations of $\Delta_0$~\cite{Singer1996From, Singer1998On} and the pairing temperature determined by the onset of reduction in $\chi_m$~\cite{Paiva2010Fermions,Fontenele2022Two}. In the manuscript, we take the mean field $T^*$ to approximately describe the pairing onset temperature. 

The many-body pairing energy scales can be compared to the two-body bound state energy $E_b$ in a lattice~(Fig.~\ref{fig:Fig_BoundState}(b)), determined by~\cite{Wortis1963Bound,Salasnich2012Pair}
\begin{equation}
    \frac{1}{U} = \frac{1}{\Omega} \sum_k \frac{1}{E_b + 2 \epsilon_k}.
\end{equation}
Here $\Omega$ is the area of the system, $\epsilon_k = 4t - 2t \cos(k_x)- 2t \cos(k_y)$, and $E_b$ is by convention positive. The energy is linear in $U/t$ for large interactions, $E_b \to U-8t$.
An exact solution for $E_b$ is given by the implicit equation
\begin{equation}
    \frac{1}{U} = \frac{2}{\pi} \frac{1}{E_b + 8t}K\left( \frac{8t}{E_b + 8t}\right),
\end{equation}
where $K(k)$ is the complete elliptic integral of the first kind of modulus $k$, $K(k) = \int_0^{\pi/2} d\theta (1- k^2 \sin^2\theta)^{-1/2}$~(see Appendix D of Ref.~\cite{Wortis1963Bound}). The limiting behavior at small $U/t$ is $E_b \approx 64 t\exp(-8\pi t/U)$, obtained from $K(k){\approx}\ln(4/\sqrt{1-k^2})$ near $k{\approx}1$.  

The mean field pairing gap $\Delta_0$ is strongly enhanced in two dimensions compared to $E_b$, and in the bulk is given by $\Delta_0{=}\sqrt{2E_bE_F}$~\cite{Miyake1983Fermi,Randeria1990Superconductivity}. 

\subsection{Fluctuation-dissipation theorem}
The grand canonical partition function $Z$ at temperature $T$ is 
$Z = \text{Tr}[ e^{- \beta \hat{H}} ]$, where $\beta = 1/T$. The density $n$ and magnetization $m$ can be written as first derivatives of the grand potential $F=-(T/\Omega)\ln{Z}$,
\begin{align}
    n &= -\frac{\partial F }{\partial \mu} =  \frac{\Tr  \hat{n}_i e^{-\beta \hat{H}}}{Z}, \\
    m &= -\frac{\partial F}{\partial h} =  \frac{\Tr  \hat{m}_i e^{-\beta \hat{H}}}{Z}.
\end{align}

\begin{figure}[t]
	\centering
	\includegraphics[width=3.5 in]{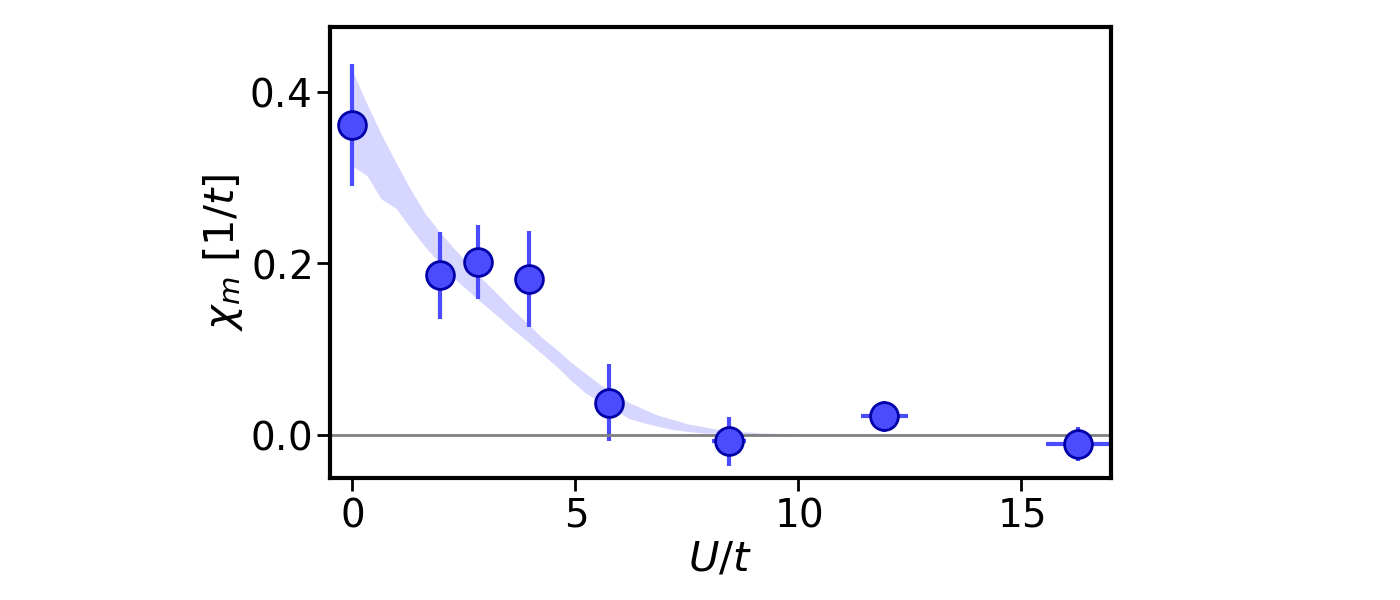}
	\caption{Uniform magnetic susceptibility $\chi_{m}{=} \partial m/\partial h$ vs.~$U/t$, obtained via the magnetization fluctuation-dissipation theorem from the total magnetization and density fluctuations, and the density susceptibility $\chi_{n}{=}\partial n / \partial \mu$. Shading shows simulations at $n{=}0.85$ and $T/t{=}0.3$ to $0.4$~\cite{Varney2009Quantum}.
	}
	\label{fig:Fig_SpinSusceptibility}
\end{figure}

The susceptibilities $\partial n/\partial \mu$ and $\partial m/\partial h$ are given by second derivatives of $F$, in terms of connected correlators $\langle \hat{n}_i \hat{n}_{i+\delta}\rangle_c  =\langle \hat{n}_i \hat{n}_{i+\delta}\rangle - \langle \hat{n}_i\rangle \langle \hat{n}_{i+\delta}\rangle$,
\begin{align}
    \frac{\partial n}{\partial \mu} &= \beta \sum_{\vec{\delta}} \langle \hat{n}_i \hat{n}_{i+\delta}\rangle_c, \\
    \frac{\partial m}{\partial h} &= \beta \sum_{\vec{\delta}} \langle \hat{m}_i \hat{m}_{i+\delta}\rangle_c.
    \label{eqn:FluctDissEqn}
\end{align}

The susceptibility to a spatially-varying perturbation is also provided by correlations. Allowing the chemical potential in Eqn.~(\ref{eqn:FHGrandCanonicalHamiltonian}) to vary as $\mu_j =  \mu_0 {+} \Delta \mu \,\text{cos} (\vec{k}\cdot \vec{x}_j) $, and taking the ratio  $\delta n_i/ \delta \mu_i$ at $\vec{x}_i=0$ gives the static susceptibility at finite wavelength, denoted as $\chi_{n}(\vec{k})$, 
\begin{align}
    \chi_{n}(\vec{k})&=
    \frac{\partial n}{\partial \mu}(\vec{k}) = \beta \sum_{\vec{\delta}}\langle \hat{n}_i \hat{n}_{i+\delta}\rangle_c \cos (\vec{k}\cdot \vec{\delta}) \\
    \chi_{m}(\vec{k})&=
    \frac{\partial m}{\partial h}(\vec{k}) = \beta \sum_{\vec{\delta}} \langle \hat{m}_{i} \hat{m}_{i+\delta}\rangle_c \cos (\vec{k}\cdot \vec{\delta}).
    \label{eqn:FluctDissEqnFiniteWavelength}
\end{align}
These relations do not depend on the specific form of $\hat{H}$ beyond the coupling to $\mu$ and $h$.

\begin{figure}[!t]
	\centering
	\includegraphics[width=3.5 in]{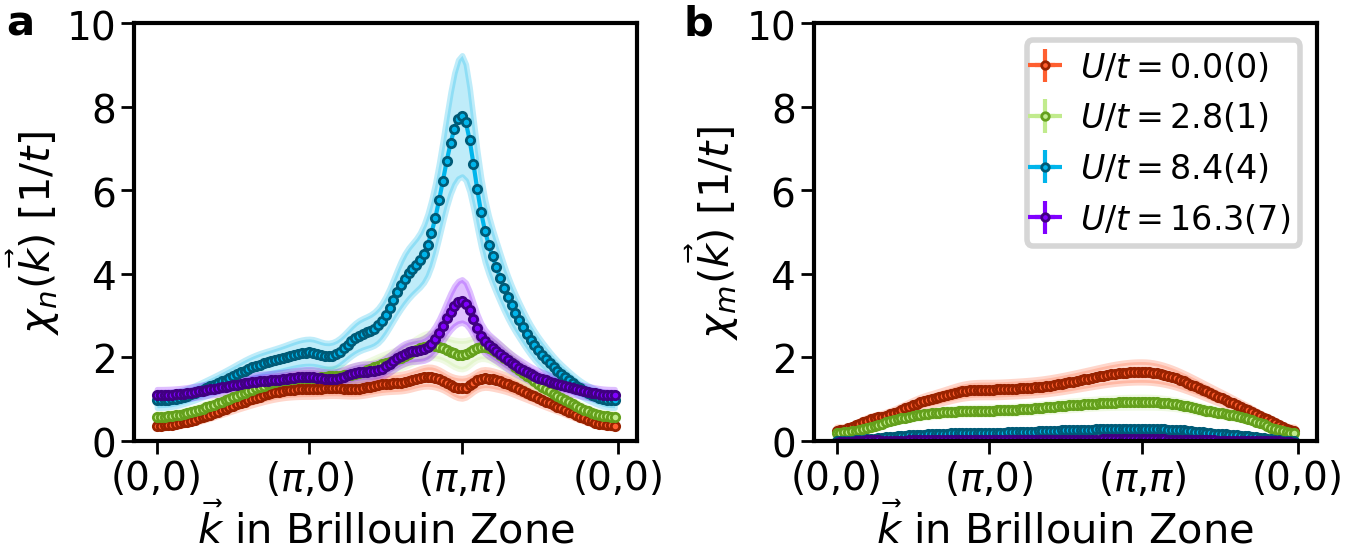}
	\caption{Measured susceptibilities of density~(a) and magnetization~(b) in a trace of $\vec{k}$ through the Brillouin zone, for various $U/t$~(see legend). Data is obtained via the fluctuation-dissipation theorem. Shading denotes error bars. 
	}
	\label{fig:Fig_Susceptibilities_vsk}
\end{figure}

Fig.~\ref{fig:Fig_SpinSusceptibility} shows the measured $\chi_{m}{=} \partial m/\partial h$ vs.~$U/t$, obtained from Fig.~\ref{fig:Fig_SpinFluctuationSummary}(c) and Fig.~\ref{fig:Fig_ChargeFluctuationsSummary}(d). 
The measured density and magnetic susceptibilities vs.~$\vec{k}$ at a few interactions $U/t$ are shown in Fig.~\ref{fig:Fig_Susceptibilities_vsk}. 

\subsection{Spin and density imaging of dense atomic clouds}

Simultaneous imaging of spin and charge is performed by Raman sideband cooling~\cite{Cheuk2015Quantum} in a bilayer optical lattice~\cite{Hartke2020Doublon}.
The procedure consists of three steps: atoms are frozen in their respective lattice sites in a single-layer system, a magnetic field gradient is applied as the system is split into a bilayer lattice, thereby mapping spin information to spatial location, and then each layer of the bilayer lattice is separately sequentially imaged, while the other layer is kept dark through interference effects. 

After preparing a system, we first ramp the lattices in $250\,\mu$s to a depth of $30\,E_R$, and then to $100\,E_R$ in an additional $2.5\,$ms, preventing tunneling. Each site may contain no atoms, a single atom of one spin type, or two atoms of different spin type. The magnetic field is typically near $203\,$G, above the Feshbach resonance of the two hyperfine states of ${}^{40}$K at $202.1\,$G. 

\subsubsection{Stern-Gerlach bilayer mapping}
We next use a Stern-Gerlach procedure to map the spin information to the bilayer location before imaging. 
In order maximize the fidelity of this mapping, we first transfer one of the spins to a state with large, opposing magnetic moment compared to the other spin. Specifically, state $\ket{1}{\equiv}\ket{F{=}9/2, m_F{=}{-}9/2}$ is converted to state $\ket{18}{\equiv}\ket{F{=}7/2, m_F{=}{-}7/2}$ using an RF Landau Zener sweep, while state $\ket{2}{\equiv}\ket{F{=}9/2, m_F{=}{-}7/2}$ is unaltered. 
However, for doubly-occupied lattice sites, transferring one of the atoms in a pair to the upper hyperfine manifold $F{=}7/2$ results in rapid spin-changing collisions and loss of that pair. Therefore, this transfer from $\ket{1}$ to $\ket{18}$ is performed only on isolated atoms by first ramping the magnetic field from ${\sim} 203\,$G to $195\,$G, where atom pairs form tightly-bound molecules with a large binding energy that are not resonantly coupled by the RF pulse. 

After transferring the isolated atoms in state $\ket{1}$ to state $\ket{18}$, the magnetic field is ramped to ${\sim}250\,$G where the atom pairs~(in state $\ket{1}$ and $\ket{2}$) now experience the background repulsive interaction of ${}^{40}$K, which produces an energy shift of approximately $6\,$kHz at $100\,E_R$ lattice depth. 
A magnetic field gradient of ${\sim} 100\,\text{G/cm}$ is applied in the out-of-plane direction, corresponding to a spin-dependent energy offset of magnitude ${\sim}6\,$kHz between the two wells for each spin state. This energy offset has the same sign for state $\ket{1}$ and state $\ket{2}$, and opposite sign for state $\ket{18}$. Isolated atoms in state $\ket{2}$ or $\ket{18}$ are therefore forced in opposite directions. In contrast, atom pairs on a lattice site are subject to a force to the same direction, though this force is ultimately overwhelmed by repulsive interactions. 

Each lattice site is then adiabatically separated out-of-plane into a bilayer lattice in $100\,$ms~\cite{Hartke2020Doublon,Hartke2022Quantum}. Within a final energy detuning window of ${\sim}6\,$kHz in the bilayer double well system, the magnetic field gradient robustly separates isolated atoms based on their spin, while the repulsive interactions separate atom pairs. 

Finally, the magnetic field gradient is removed, a quantization magnetic field of $4.2\,$G is applied along the $x$-axis for imaging, and the lattice depths are increased to $1000\,E_R$~\cite{Cheuk2015Quantum}.

\subsubsection{Bilayer-selective Raman sideband imaging}
Subsequent to the bilayer Stern-Gerlach mapping, no lattice site is occupied by more than one atom, and therefore no atom can be lost due to light-assisted collisions in fluorescence imaging.
Raman sideband cooling light is then applied to cause layer-selective atomic fluorescence~\cite{Hartke2020Doublon}. The imaging light consists of $F{=}7/2$ pumping light and repumping $F{=}9/2$ light close to the D1 transition, and two Raman beams close to the D2 transition, as described previously in Ref.~\cite{Cheuk2015Quantum}. The $F{=}7/2$ pumping light is circularly polarized and propagates along the $x$-axis, illuminating both of the bilayer lattice layers with similar intensity.

Differential imaging of the two layers is achieved by manipulating the geometry of the two Raman beams~(propagating along the $x$-axis and $y$-axis, respectively) and the repumping $F{=}9/2$ light~(co-propagating through the same fiber with the Raman light along the $y$-axis) to place one layer simultaneously at an interference node of all three of these light sources. Each of the two Raman beams is directly counter-propagating to the incoming path of one of the optical lattice beams, and has identical polarization~(in the $x{-}y$~plane), and therefore forms a high contrast interference lattice upon reflection from the microscope substrate. This interference lattice is precisely referenced to the position of the bilayer optical lattice, since both are set by the reflection from the microscope substrate. However, the interference node position of each Raman beam can be tuned by changing its angle of incidence on the microscope using a motorized glassplate in a Fourier plane. To selectively image one layer of the bilayer system, we set the node of the Raman beams~(and thus also the repumping $F{=}9/2$ light) to be located at the other layer. The layer located at a node is then only subject to the circularly polarized $F{=}7/2$ pumping light, which illuminates both layers. Atoms in this layer are quickly pumped to a dark state of the $F{=}7/2$ pumping light, where they remain without further scattering. 

In an experiment, we first image the upper layer~(denoted as spin $\uparrow$, atoms in state $\ket{F{=}9/2, m_F{=}{-}7/2}$) by placing the lower layer~(denoted as spin $\downarrow$, atoms currently in state $\ket{F{=}7/2, m_F{=}{-}7/2}$, originally in state $\ket{F{=}9/2, m_F{=}{-}9/2}$) at a node of the imaging light. Illumination is paused after ${\sim}2\,$s, and the node is moved to the upper layer in ${\sim}0.5\,$s, before a second image of the lower layer is collected for ${\sim}2\,$s. The background fluorescence of the layer placed at the node is not detectable in a given image. 
Typical loss rates during an image are 7(2)\%~(8(2)\%) while fluorescing, and 7(2)\%~(6(2)\%) when placed at the imaging node, for the upper~(lower) layer. Typical hopping rates (${<}1\%$) and misidentification rates (${\sim}2(1)\%$) are small, and are neglected in imaging loss corrections. Comparable loss rates are observed in clouds with large or small doublon number, indicating a lack of inter-layer correlated loss. 
Multiple sequential images of the same cloud are taken during each experimental run to directly measure all loss and hopping rates in various configurations. 

Finally, we note that the microscope position is not adjusted during imaging because the two layers of the bilayer system are separated by $532\,$nm along the imaging axis, less than the optical wavelength $767\,$nm of the light emitted by each atom. Therefore atoms in both layers of the bilayer lattice remain within the diffraction-limited focus of the microscope objective~\cite{Hartke2020Doublon, Cheuk2015Quantum}. To reconstruct the full atomic density, binned images of the lattice occupation in each layer are simply combined. 

\subsection{Extraction of densities and correlations}

For clarity, we here summarize a few aspects of data processing: 
\begin{itemize}
    \item  Data and error bars at each $U/t$ are obtained from bootstrapping greater than 50 images of atomic cloud, using the central region of radius 10 sites.
    \item Images are post-selected for globally spin balanced systems~(typically within $\pm$4\% total imbalance $(N_\uparrow {-} N_\downarrow)/(N_\uparrow {+} N_\downarrow)$ in the entire atomic cloud).
    \item Loss of atoms during imaging is accounted for in reported densities and correlations~(see below). 
    \item A calibrated uniform offset is applied to two-point correlations to account for global density fluctuations and the spatial variation of density within the sample area~(for details see below).
    \item Magnetization fluctuation sums include correlations out to $|\vec{\delta}|{=}4 {\pm} 0.5$, with randomization during bootstrapping to reduce sensitivity to the cutoff.
    Density correlation sums $\sum_{\vec{\delta}}\langle \hat{n}_i \hat{n}_{i+\delta}\rangle_c \text{cos} (\vec{k}\cdot \vec{\delta})$ are obtained from density correlations after smoothing data beyond 2.5 sites via an exponential fit to the rectified density correlations $\sum_{\vec{\delta}}\langle \hat{n}_i \hat{n}_{i+\delta}\rangle_c (-1)^{\delta_x + \delta_y}$ vs.~$|\vec{\delta}|$, which we find to be a good description in all data. The same procedure is used for inter-spin correlation sums. 
\end{itemize}

\subsubsection{Corrections for imaging loss}
A fraction of atoms are lost during the process of scattering light for atomic detection, necessitating loss correction to report estimates of true values.
We use the variable $\tilde{n}_{i\sigma}$ to denote the observed density of spin $\sigma$ on site $i$. 
With a total loss rate $l_{\sigma}$ of spin $\sigma$ before imaging is complete, the inferred true density is $\langle \hat{n}_{i\sigma}\rangle = \langle \tilde{n}_{i\sigma} \rangle /(1-l_{\sigma})$. Connected correlations between separated lattice sites $i$ and $j$ with $i\neq j$ are corrected as
\begin{equation}
\langle \hat{n}_{i\sigma} \hat{n}_{j\sigma'}\rangle_c =\frac{ \langle \tilde{n}_{i\sigma} \tilde{n}_{j\sigma'} \rangle_c }{(1-l_{\sigma})(1-l_{\sigma '})} \hspace{4mm} (i\neq j). 
\end{equation}
When measuring the same species on the same lattice site, a correction by only one factor of $(1-l_{\sigma})$ is required, i.e.~$\langle \hat{n}_{i\sigma} \hat{n}_{i\sigma}\rangle = \langle  \tilde{n}_{i\sigma} \rangle/(1-l_{\sigma})$. 
This correction is easily generalized to apply to three-point correlations, as in Fig.~\ref{fig:Fig_Polaron}.
Here, it should be understood that the stated spin index refers to a physical layer after the Stern-Gerlach mapping. Loss $l_\downarrow$ therefore occurs during the first two images, while loss $l_\uparrow$ occurs only during the first image. 

\subsubsection{Corrections for atom number variation}
Most cold atom experiments possess inherent atom number fluctuations which produce offsets to measured correlations regardless of underlying physics. 
Generally, given two variables which possess no physical correlations, such as $\hat{n}_{i \uparrow}$ and $\hat{n}_{i + \delta \uparrow}$ with $|\vec{\delta}| \gg 1$~(so that distance implies a lack of correlation), the measured connected correlator $\langle \hat{n}_{i \uparrow} \hat{n}_{i + \delta \uparrow} \rangle_c = \langle ( \hat{n}_{i \uparrow}{-}\langle  \hat{n}_{i \uparrow} \rangle )(\hat{n}_{i + \delta \uparrow } {-}\langle \hat{n}_{i + \delta \uparrow}  \rangle)\rangle $ will still be nonzero due to experimental fluctuations in the average density of $\uparrow$ atoms within a large region. Denote the density of spin $\uparrow$ atoms in a large box within a single experiment as $\hat{n}_{\uparrow \boxdot}$. One can show that $\langle \hat{n}_{i \uparrow} \hat{n}_{i + \delta \uparrow} \rangle_c$ will equal the variance of $\hat{n}_{\uparrow \boxdot}$ over many repeated experiments, $\sigma^2_{\uparrow \boxdot}=\langle ( \hat{n}_{\uparrow \boxdot}{-}\langle  \hat{n}_{\uparrow \boxdot} \rangle )(\hat{n}_{\uparrow \boxdot} {-}\langle \hat{n}_{\uparrow \boxdot}  \rangle)\rangle$. Likewise, correlations between fluctuations of spin $\uparrow$ and spin $\downarrow$ will cause a uniform offset to $\langle \hat{n}_{i \uparrow} \hat{n}_{i + \delta \downarrow} \rangle_c$ equal to the cross correlation $\sigma^2_{\uparrow,\downarrow \boxdot}=\langle ( \hat{n}_{\uparrow \boxdot}{-}\langle  \hat{n}_{\uparrow \boxdot} \rangle )(\hat{n}_{\downarrow \boxdot} {-}\langle \hat{n}_{\downarrow \boxdot}  \rangle)\rangle$.
Such error is typically irrelevant in experiments, but is magnified by summing long range two-dimensional correlations maps, as in Fig.~\ref{fig:Fig_SpinFluctuationSummary} and  Fig.~\ref{fig:Fig_ChargeFluctuationsSummary}. To account for such systematic biases, we directly measure within each dataset at each $U/t$ the effective variances and cross correlation  $\sigma^2_{\uparrow \boxdot}$, $\sigma^2_{\downarrow \boxdot}$, and $\sigma^2_{\uparrow,\downarrow \boxdot}$, which are then subtracted from $\langle \hat{n}_{i \uparrow} \hat{n}_{i + \delta \uparrow} \rangle_c$, $\langle \hat{n}_{i \downarrow} \hat{n}_{i + \delta \downarrow} \rangle_c$, and $\langle \hat{n}_{i \uparrow} \hat{n}_{i + \delta \downarrow} \rangle_c$, respectively.

We account for two sources of fluctuations in measuring $\sigma^2_{\uparrow \boxdot}$, $\sigma^2_{\downarrow \boxdot}$, and $\sigma^2_{\uparrow,\downarrow \boxdot}$. 
One contribution is provided by the image-to-image fluctuations in the average densities $n_{\uparrow, {\rm img}}$ and $n_{\downarrow, {\rm img}}$ within the entire central region of the atomic cloud.
A second contribution arises from the spatial variation of the densities $n_{\uparrow}(r)$ and $n_{\downarrow}(r)$ within that region after averaging together all images. For small fluctuations, these two effects are uncorrelated and can be directly summed, i.e.~$\sigma^2_{\uparrow \boxdot} = \sigma^2_{n_{\uparrow}, {\rm img}} + \sigma^2_{n_{\uparrow}(r)}$. 

Typical resulting corrections to each measured correlator are of order $6(2){\times} 10^{-4}$, with similar effects from spatial variation and total number fluctuations. Data are corrected throughout the paper, excluding the data in Fig.~\ref{fig:Fig_Polaron} because it does not include correlation sums, only individual correlators. 
Magnetization-magnetization correlations are essentially unaffected by this correction, since the intra-spin and inter-spin corrections are comparable, and cancel. 
As a final note, this correction procedure assumes $\hat{n}_{i \sigma}$ and $\hat{n}_{i + \delta \sigma'}$ are uncorrelated, regardless of their spatial displacement $\vec{\delta}$, which is not guaranteed for small $|\vec{\delta}|$. The correction to some short-range correlators will therefore be incorrect. However, this correction controls bias in long range correlation sums, and has minimal effect on individual correlators.

\subsection{Quantum Monte Carlo simulation of magnetic fluctuations}
Numerical simulations of the magnetic fluctuations are performed using the Quantum Electron Simulation Toolbox (QUEST) Fortran package~\cite{Varney2009Quantum}. For simulation results shown in Fig.~\ref{fig:Fig_SpinFluctuationSummary} and Fig.~\ref{fig:Fig_SpinSusceptibility}, we employ a homogeneous $8{\times}8$ site lattice. The simulation starts with $3,000$ warmup sweeps followed by $7,000$ measurement sweeps. The number of imaginary time slices is set to 60 to achieve reliable results at low temperatures. The total magnetic fluctuations reported in the main text are obtained by taking the direct sum of the magnetization correlators up to a displacement of 4 lattice sites.

\end{document}